\documentclass{aa}

\usepackage{txfonts}
\usepackage{longtable}
\usepackage{xcolor}
\usepackage{graphicx} 
\usepackage{comment}
\usepackage[normalem]{ulem}
\usepackage[tbtags,fleqn]{amsmath}

\begin{document}

\newcommand{\kms}{km\,s$^{-1}$}
\newcommand{\Msun}{\ensuremath{M_\odot}}

\title{Projection effects in barred galaxies causing wrong interpretation of radial flows}

\author{E. Salibur\inst{\ref{inst1}} \and A. Hall\'e\inst{\ref{inst1}} \and F. Combes\inst{\ref{inst1},\ref{inst2}}}
\institute{Observatoire de Paris, LUX, CNRS, PSL University, Sorbonne University, 75014, Paris \label{inst1} 
\and Coll\`ege de France, 11 Pl. Marcelin Berthelot, 75005 Paris, France \label{inst2}}

\abstract{Galaxy disks in rotation are sometimes the site of radial flows, especially in their gas component. It is 
important to estimate the outflows, due to AGN or supernovae feedback, or inflows due to bar gravity torques. However, these radial flows may be confused with non-circular motions, which are quite frequent in the center of galaxy disks.
We use a simulated giant, barred spiral galaxy from the GalMer database to study the non-circular motions induced by the bar. Our goal is to identify the spurious radial flows that kinematics modeling algorithms can detect, assuming circular orbits for the gas. Using mock data of a strongly barred galaxy, we quantify the radial velocities computed by the \texttt{3D-Barolo} algorithm for different disk inclinations and several bar orientations in the plane of the sky: along the major and minor kinematic axes and at $\pm 45^\circ$ from them. Our results show that projection effects cause kinematics modeling algorithms to confuse the radial component of velocity due to elliptical orbits with significant radial flows with mean values up to 92~\kms\, within the bar region. The computed rotation curve is also wrongly estimated inside the bar region, by as much as 150~\kms for the highest inclination.}


\maketitle

\section{Introduction} 

More than two-thirds of the local universe spiral galaxies are barred \citep{1991rc3..book.....D, 1999Ap&SS.269..427E}. Barred spiral galaxies are typically classified into two categories: SAB (moderately barred) and SB (strongly barred), depending on the bar's strength.

Bars are associated with an $m=2$ instabilities
with two-arm spiral waves, producing a bi-symmetric gravitational potential \citep{Buta1996}. They represent a long-lived, quasi-stationary density wave arising from the superposition of leading waves, which transfer angular momentum inward, and trailing waves, which transfer angular momentum outward. Bars exchange angular momentum with the outer disk and dark matter halo through resonant interactions, particularly at the corotation and outer Lindblad resonances (OLR), which allows them to grow and slow down. 
Bars induce spiral structure in the gas component, which is dissipative: gas cloud collisions prevent the gas to follow crossing 
orbits like stars \citep{Sanders1976, Buta1996}. Bars and spirals are then out of phase, and bars exert gravity torques on the gas. Inside the corotation region (CR), these torques are negative on gas weighted average, and the gas loses angular momentum to the benefit of the bar. In fact, in active galactic nuclei (AGNs), the gas angular momentum has to be reduced by 5 orders of magnitude when infalling from a typical galactocentric radius of 3~kpc to the last stable orbit, over a timescale much shorter than that of the known efficient mechanisms leading to gas accretion onto the central supermassive black hole (SMBH) \citep{2023Galax..11..120C}. Bars also exchange angular momentum with the outer disk and the dark matter halo through resonant interactions, particularly at the corotation radius and OLRs, which allows them to grow and slow down \citep{2002A&A...392...83B}.

Bars and spiral arms formation has been reproduced by $N$-body simulations at least since \cite{1970ApJ...161..903M} who obtained persistent spiral arms, in addition to a long-lasting bar, thanks to the addition of the gas component. Simulations have shown how the bar works as an angular momentum transport mechanism to fuel the central engine, and to increase the star formation rate \citep{2010MNRAS.407.1529H}. They also demonstrate how a secondary bar may decouple, to present morphologies like the 'bars-within-bars' initially proposed by \cite{1989Natur.338...45S} that could solve the lack of efficiency of large-scale bars in driving gas inward. Indeed, primary bars only drive the gas down to the inner Lindblad resonance (ILR) but not inside it.

In specific regions of barred galaxies, the gas can be trapped in rings at resonances, when there is a rational multiple between motion frequencies of the particles and the bar wave, in the bar rotating frame (see \cite{1989A&ARv...1..261C} for a review on stellar orbits in barred galaxies). When the orbital frequencies satisfy the condition for the $m=2$ Lindblad resonances (for a bar wave), i.e. when $\Omega\pm\kappa/2=\Omega_b$ where $\Omega$ is the circular frequency, $\kappa$ the epicyclic frequency, and $\Omega_b$ the bar pattern speed, the orbits close after two radial oscillations \citep{Buta1996} in the bar rotating frame. Inside corotation, the ILR, associated with a negative sign in the previous equation, plays a role in damping waves that might prevent bar formation, or weaken the bar. Stellar orbits orientation alternates between parallel ($x_1$ orbits) and perpendicular ($x_2$ orbits) to the bar at each resonance \citep{1980A&A....92...33C, 1992MNRAS.259..328A}. These orbits support the bar structure. At other radii than the resonance ones, the inclination of the gas orbits varies, they do not change drastically, hence the spirals \citep{2014A&A...565A..97C}.

\cite{2014A&A...565A..97C} were the first to observe a trailing spiral structure driving gas inward within the ILR ring of a nuclear bar in NGC~1566, showing what kind of morphological features allow the gas to fuel the central SMBH. In fact, negative azimuthally-averaged and gas-weighted torques have been computed in the central 300 pc, removing angular momentum from the gas under the SMBH’s influence. At a resonance, the sign of the torque exerted by the bar on the gas changes, causing gas to accumulate in rings. Inside the ILR, the central mass concentration can modify the potential, changing the precession rate and reversing the torque sign again, so only trailing spirals persist, funneling gas inward toward the nucleus \citep{2023Galax..11..120C}. This also weakens the bar \citep{2002A&A...392...83B}.

As shown by \cite{1982MNRAS.198..517B, 1984MNRAS.207....9P, 1984MNRAS.210..547P}, bars affect gas dynamics and particularly induce non-circular motions because of their non-axisymmetric nature. Interpreting radial motions in barred galaxies is not trivial. Projection effects primarily affect the gas around the bar. The radial component of the gas velocity due to elliptical orbits induced by the barred potential, can be mistaken for radial flows. Projection effects in barred galaxies have been addressed by \cite{2000AstL...26..565M}. Using 2D models, they showed that for highly inclined galaxies, the bar could lead to a false identification of a warped disk. The projection is due to the intrinsic orbital asymmetry of elliptical orbits in barred potentials. A wrong interpretation or estimation of radial velocities from kinematics models assuming axisymmetry biases inflow or outflow rate computations.

Observational programs such as NUclei of GAlaxies (NUGA) and the Galaxy Activity, Torus, and Outflow Survey (GATOS) investigate the gas dynamics and activity in the close environment of AGN, at the sub-kpc scale. While NUGA focused on circumnuclear molecular gas and gravitational torques, GATOS targets ionized and molecular outflows on similar scales, probing the interplay between inflow and feedback. The computation of gravity torques from the bar have revealed negative gas-weighted averaged torques inside CR, leading to gas inflow toward the nucleus \citep{2011A&A...527A..92C}. These systems often display embedded structures, such as 'bars-within-bars' configurations \citep{2008arXiv0811.1971C}. In several galaxies from the GATOS survey, ionized gas outflows have been observed and characterized \citep{2024A&A...689A.263D, 2024ApJ...974..195Z, 2024A&A...690A.350H}. Ionized outflows have a complex geometry and not necessarily match the molecular outflow \citep{2024A&A...689A.263D}. They could be triggered by radio jets \citep{2024ApJ...974..195Z} and cause positive feedback inducing star formation in the interstellar medium of the galaxy \citep{2024A&A...690A.350H}. In order to properly quantify these radial flows, it is important to distinguish and separate in the gas kinematics what is due to non-circular motions, from the actual flows. \cite{2016A&A...594A..86R} have shown with simulated mock galaxies, analyzed with the algorithms \texttt{ROTCUR} and \texttt{DiskFit}, that the gas kinematics can be strongly biased, and a true rotation curve difficult to obtain.

In this work, we use a simulated barred galaxy from the GalMer database \citep{2010A&A...518A..61C} as this allows us to control the true kinematics and isolate projection effects from intrinsic motions, providing a benchmark for interpreting observational modeling with the 3D-Based Analysis of Rotating Objects from Line Observations (\texttt{3D-Barolo}) software \citep{2015MNRAS.451.3021D, 2021ApJ...923..220D}. 

This paper is organized as follows. In Sect.~\ref{sec:data} we present the mock galaxy used in this study based on GalMer data. The parameters we input in \texttt{3D-Barolo} and our staged approach are detailed in Sect.~\ref{sec:methods}. We modeled the gas kinematics for different bar orientations: the models with the bar along the kinematic axes are presented in Sect.~\ref{sec:symaxes} and with the bar at $\pm 45^\circ$ from the kinematic axes in Sect.~\ref{sec:bar45}. Our different results are discussed and compared especially with works using similar methods on observations in Sect.~\ref{sec:discussion}. The main conclusions of our study are summarized in Sect.~\ref{sec:conclu}.

\section{Data}\label{sec:data}

We based our study on a gas-rich giant spiral barred galaxy (gSb) from the GalMer database. This database has been built using smoothed-particle hydrodynamics (SPH) simulations to study the impact of intergalactic interactions on galaxy evolution, but it contains isolated galaxy runs as control samples. For the purpose of our analysis we selected one of these isolated galaxy runs. We chose one with 160 000 particles (run \#1030). We wanted well-defined spiral arms and a strong bar so as in \cite{2016A&A...594A..86R} we analyzed the snapshot at $T=500$ Myrs. From now on, the model will be referred to as gSb. The properties of gSb (mass of the different components, scale lengths and number of gas particles) are given in Table~\ref{tab:gSb_prop}.

\begin{table}
\caption{Fundamental parameters for gSb.}
\label{tab:gSb_prop}
    \centering
    \begin{tabular}{cc}
    \hline
    \hline
    Parameter & Value \\
    \hline
        $M_{\rm{B}}\ [10^9 $\Msun] & 11.5\\
        $M_{\rm{H}}\ [10^9$\Msun] & 172.5\\
        $M_{*}\ [10^9$\Msun] & 46\\
        $M_{\rm{g}}\ [10^9$\Msun] & 9.2\\
        $r_{\rm{B}}$ [kpc] & 1\\
        $r_{\rm{H}}$ [kpc] & 12\\
        $a_*$ [kpc] & 5\\
        $h_*$ [kpc] & 0.5\\
        $a_{\rm{g}}$ [kpc] & 6\\
        $h_{\rm{g}}$ [kpc] & 0.2\\
        $N_{\rm{part}}$ & 160 000\\
        \hline
    \end{tabular}
    \tablebib{\citet{2010A&A...518A..61C}}
\end{table}

The dark matter halo and stellar bulge of the gSb model were represented initially by Plummer spheres \citep[p.~42]{1987gady.book.....B}. The corresponding density profiles are:
\begin{equation}\label{eq:plummer}
    \rho_{H,B}(r) = \frac{3M_{H,B}}{4\pi r_{H,B}^3}
    \left(1+\frac{r^2}{r_{H,B}^2}\right)^{-5/2},
\end{equation}
where $M_H$ and $r_H$ are the characteristic mass and radius of the dark matter halo, and $M_B$ and $r_B$ those of the bulge (see Table~\ref{tab:gSb_prop}). The corresponding Plummer potentials are:

\begin{equation}\label{eq:plummer_halo}
    \Phi_{H,B}(r) = -\,\frac{GM_{H,B}}{\sqrt{r^2 + r_{H,B}^2}}.
\end{equation}

The stellar and gaseous disks follow initially a Miyamoto-Nagai density profile \citep[p.~44]{1987gady.book.....B}:
\begin{equation}
\begin{split}
    \rho_d(R,z) = \frac{h^2 M}{4\pi}\,
    &\frac{aR^2 + (a + 3\sqrt{z^2 + h^2})(a + \sqrt{z^2 + h^2})^2}
    {[a^2 + (a + \sqrt{z^2 + h^2})^2]^{5/2}(z^2 + h^2)^{3/2}},
\end{split}
\end{equation}
where $M$ is the disk mass ($M_g$ or $M_*$), and $a$ ($a_g$ or $a_*$) and $h$ ($h_g$ or $h_*$) are the radial and vertical scale lengths (see Table~\ref{tab:gSb_prop}).  
The associated potential is:

\begin{equation}\label{eq:miyamoto}
    \Phi_d(R,z) = -\,\frac{GM}{\sqrt{R^2 + [a + \sqrt{z^2 + h^2}]^2}}.
\end{equation}

From the GalMer website\footnote{\url{http://galmer.obspm.fr}}, we obtained masses, positions and velocities of gas particles. We added the same particles at opposite positions $(-x, -y, -z)$ and opposite velocities $(-v_x,-v_y,-v_z)$ to the galaxy in order to reduce morphological asymmetries, as the different configurations studied are created by rotation, and we want them to differ only by the bar's orientation. Then, spectral cubes are created only with the gas particles with each pixel measuring 200~pc $\times$ 200~pc and using a spectral resolution of 12~\kms. The pixel size value represents the minimum softening length used in GalMer \citep{2010A&A...518A..61C}. The bar is about 5~kpc long, which is equivalent to $\sim 25$ pixels. Initially, the galactic disk is in the $x-y$ plane and the line-of-sight (LoS) is in the $z$ direction. We create one cube every 10$^\circ$ for an inclination angle $i_{\rm{true}}\in[30^\circ,80^\circ]$, the interval recommended by \cite{2021ApJ...923..220D}. In order to study symmetrical configurations with respect to the LoS, we vary the inclination orientation which is equivalent to changing the nearest/farthest side. To reproduce observations we smooth the obtained cube using a Gaussian beam with a full width at half maximum (FWHM) of 0.3~kpc (corresponding to 1$\arcsec$=1~kpc for a galaxy at a distance of $D=206.3$ Mpc). This FWHM corresponds to a maximum baseline of $B_{\rm max}=724$ m at $\nu = 350$ GHz, consistent with ALMA CO(3–2) observations in a compact configuration. As an example, this typical emission line was used by \cite{2019A&A...632A..33A} to observe a trailing spiral inside the ILR ring of the bar in NGC 613.

\section{Methods}\label{sec:methods}

We model the gas kinematics with the \texttt{3D-Barolo} algorithm \citep{2015MNRAS.451.3021D, 2021ApJ...923..220D}. It fits emission-line datacubes using 3D tilted-rings: it divides the galactic disk in rings and computes geometrical and kinematic parameters for each ring depending on the free parameters chosen by the user. In particular, the algorithm has been used to identify and characterize radial flows in barred spiral galaxies in many studies discussed in Sect.~\ref{sec:discussion}.

The program models the gas LoS velocity $V_{\rm{LoS}}$ as
\begin{equation}
    V_{\rm{LoS}}(R,\theta)=V_{\rm{sys}}+\left(V_{\rm{rot}}(R)\cos{\theta}+V_{\rm{rad}}(R)\sin{\theta}\right)\sin{i(R)},
\end{equation}
with the galactocentric radius $R$, the azimuthal angle $\theta$, the systemic velocity $V_{\rm{sys}}$ of the galaxy center fixed to zero during all steps, the tangential velocity $V_{\rm{rot}}$, the radial component of the velocity $V_{\rm{rad}}$ and the inclination angle $i$. There is no ambiguity between radial and tangential contributions to the LoS velocity only along the kinematic axes. Along the major axis (green line on the diagrams, right column of Fig.~\ref{fig:configs}), we can only measure the tangential velocity and along the minor axis (red line on the diagrams, right column of Fig.~\ref{fig:configs}), only the radial velocity.

We ran \texttt{3D-Barolo} on the final cubes (symmetrized and smoothed) using a three-step staged approach to reduce degeneracies between the geometrical parameters and the radial velocities. When initial values are not provided by the user, \texttt{3D-Barolo} estimates them automatically.

During the first step, we do not input the initial values of the parameters to be fitted. In the first two steps, we assume an axisymmetric disk and circular velocity, neglecting the bar as it would typically be done for real observations. The geometrical parameters: the inclination angle $i$ and/or (depending on the case) the position angle (PA) $\phi$ and the kinematic parameters: the tangential velocity $V_{\rm rot}$ and velocity dispersion $V_{\rm disp}$ are fitted without radial motion ($V_{\rm rad}=0$).

The values of $i$ and/or $\phi$ obtained in the step 1 are either median or Bézier-interpolated values which smooths ring-to-ring fluctuations. The code chooses a median value if the dispersion is lower than $3^\circ$ after the first run. The values of $V_{\rm rot}$ and $V_{\rm disp}$ from step 1 serve as initial guesses for step 2, where the geometrical parameters are always fixed and $V_{\rm rot}$ and $V_{\rm disp}$ are fitted again. Finally, in the third step, $V_{\rm rot}$ and $V_{\rm disp}$ are fixed to the values from step 2, and only $V_{\rm rad}$ is fitted. With this setup, any recovered radial component reflects the intrinsic influence of the bar rather than projection effects or a warped disk.

According to \cite{2021ApJ...923..220D}, \texttt{3D-Barolo} adopts the convention that, for clockwise (respectively counter-clockwise) rotation, positive radial velocities correspond to outflow (inflow), while negative velocities correspond to inflow (outflow). Throughout this paper, we follow the following convention: positive (negative) radial velocities are interpreted as outflow (inflow). The spiral arms are trailing, which determined the rotation sense of the gas.

We made computations at specified radii with more numerous radii closer to the center. The scale-height of the mock disk was fixed to 200 pc. The masks were created by \texttt{3D-Barolo}, using the \texttt{SEARCH} option, which identifies emission regions. The normalization was made locally (pixel by pixel) and we applied a uniform weight.

We compare the tangential velocities to the circular velocity $V_{\rm{circ}}$: the expected rotation curve for circular orbits derived from the gravitational potential $\Phi(R)$ computed from the different components contribution (Eq.~\ref{eq:plummer}--Eq.~\ref{eq:miyamoto}) at the selected snapshot $T=500$~Myr. It was derived for particles at $z<100$ pc, half of the estimated scale-height, using:
\begin{equation}\label{eq:Vcirc}
    V^2_{\rm{circ}}=R\langle\frac{\partial \Phi}{\partial R}\rangle,
\end{equation}
where the brackets represent a weighted average over a concentric ring.

\section{Kinematics modeling with the bar along the kinematic axes}\label{sec:symaxes}

\begin{figure}[ht]
    \centering
    \resizebox{\hsize}{!}{
    \includegraphics[width=0.44\textwidth]{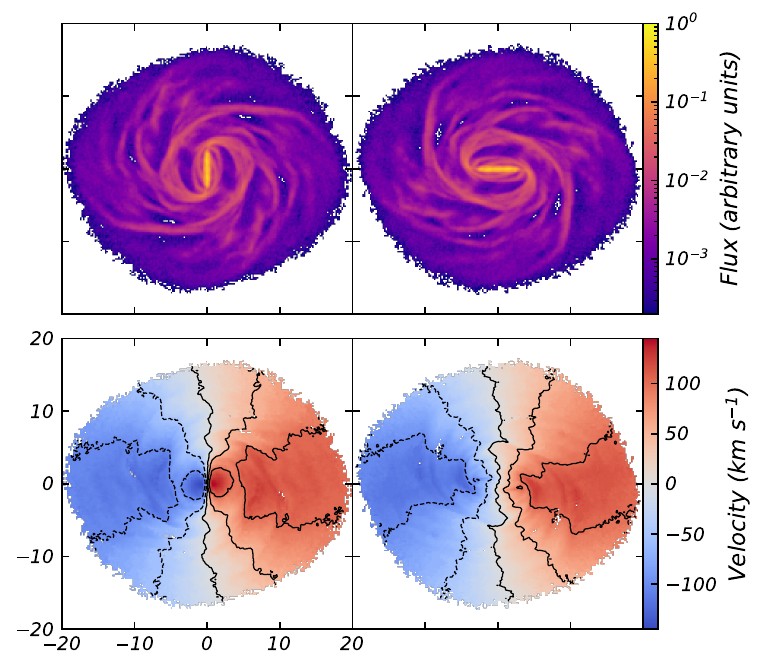}}
    \caption{The data moment maps of the gas in the central 20~kpc produced with the symmetrized gSb at the selected snapshot ($T = 500$ Myr). The pixel scale is 0.2~kpc. The point (0, 0) corresponds to the center of the bar. The gas component has been inclined by $i_{\rm{true}} = 30^\circ$ and rotated so the bar is aligned with the kinematic axes: the galaxy minor axis (left column) and major axis (right column). First row: integrated intensity maps (moment 0) in logarithmic norm. Second row: mean velocity field (moment 1) maps. The black curves show the isovelocity contours of levels [-150, -100, -50, 0, 50, 100, 150]~\kms.}
    \label{fig:bar_symaxes_moms}
\end{figure}

\begin{figure}[ht]
    \centering
    \resizebox{\hsize}{!}{
    \includegraphics[width=0.44\textwidth]{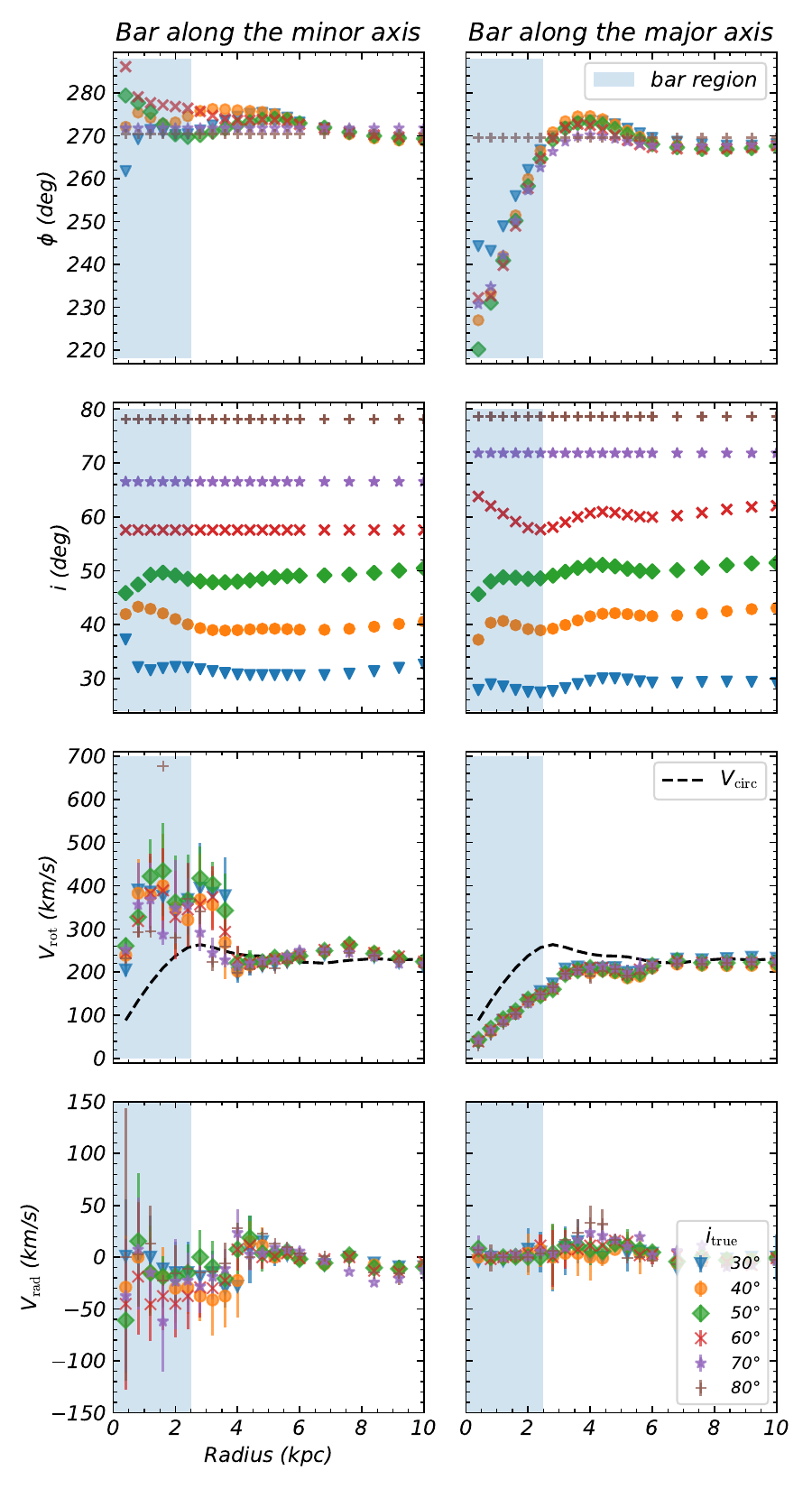}}
    \caption{Each column corresponds to one bar orientation of Fig.~\ref{fig:bar_symaxes_moms}. Left column: bar along the minor axis and right column: bar along the major axis. Parameters fitted by \texttt{3D-Barolo} with a free PA and a free inclination angle. Each marker corresponds to one 'true' inclination value $i_{\rm{true}}\in\pm[30^\circ,80^\circ]$, the one we input to create the cube. The different parameters are plotted as function of the galactocentric radius and the blue area highlights the bar region. First row: the PA obtained at the end of step 1. Second row: the inclination angles obtained at the end of step 1. Third row: rotation curves obtained after step 2. The black dashed curve represents the expected circular velocity (Eq.~\ref{eq:Vcirc}) computed from the gravitational potential (Eq.~\ref{eq:miyamoto}). Fourth row: Radial velocity with the sign convention: negative towards the center.}
    \label{fig:bar_symaxes_params}
\end{figure}

We considered two extreme bar orientations by aligning it with the major and the minor kinematic axes of the galactic disk (see moment maps in Fig.~\ref{fig:bar_symaxes_moms} for a low inclination $i_{\rm true}=30^\circ$).
This allows us to evaluate how these limiting configurations affect \texttt{3D-Barolo}'s recovery of the geometrical and kinematic parameters plotted in Fig.~\ref{fig:bar_symaxes_params}.

Two well-defined winding spiral arms connected to the bar are visible in the moment 0 map (first row of Fig.~\ref{fig:bar_symaxes_moms}) and the highest flux is along the bar which is gas-rich. This represents the most common morphology of barred spiral galaxies as 70\% of them have two arms starting at the end of the bar \citep{1982MNRAS.201.1021E}. On the velocity field map (second row of Fig.~\ref{fig:bar_symaxes_moms}), we recognize a rotation dominated disk with almost perpendicular kinematic axes.

We fit both of the geometrical parameters: the PA $\phi$ and the inclination angle $i$ and the kinematic parameters: the tangential velocity $V_{\rm rot}$, the velocity dispersion $V_{\rm disp}$ and the radial velocity $V_{\rm rad}$, plotted in Fig.~\ref{fig:bar_symaxes_params}. 

The gas streams along the bar in elongated orbits, with a velocity higher than circular at the pericenter, and lower than circular at apocenter. When the bar is along the minor kinematic axis, the higher tangential velocity component is mostly orthogonal to the major axis in the bar region so the rotation curve is higher than the circular velocity (left column, third row, panel of Fig.~\ref{fig:bar_symaxes_params}). When the bar is aligned with the major axis, the higher tangential velocity component projects along the major axis and is not observed. The recovered rotation curve is instead lower than the circular velocity (right column, third row, panel of Fig.~\ref{fig:bar_symaxes_params}). In the latter case, the bar lies along the projected axis that is least affected by inclination errors, so the rotation curve is less sensitive to inclination changes. \texttt{3D-Barolo} retrieves well those effects.

Because radial motions project onto the minor axis, the bar–driven radial component is absent in the recovered kinematics when the bar is aligned with the major axis (right column, last row, panel of Fig.~\ref{fig:bar_symaxes_params}), whereas a non-zero radial component appears when the bar is along the minor axis (left column, last row, panel of Fig.~\ref{fig:bar_symaxes_params}). These amplitudes remain lower than in the $\pm45^\circ$ configurations, explaining our choice of dedicated analysis of those orientations in Sect.~\ref{sec:bar45}.

Finally, when radial contribution is almost zero, that is when the bar is along the major axis, \texttt{3D-Barolo} cannot explain the bar effects by fitting radial velocities, so it shifts the PA by up to 55$^\circ$ within the bar region ($R<2.5~\rm kpc)$.

\section{Kinematics modeling with the bar at $\pm45^\circ$ from the kinematic axes}\label{sec:bar45}

\begin{figure*}[ht]
    \centering
    \includegraphics[width=0.8\textwidth]{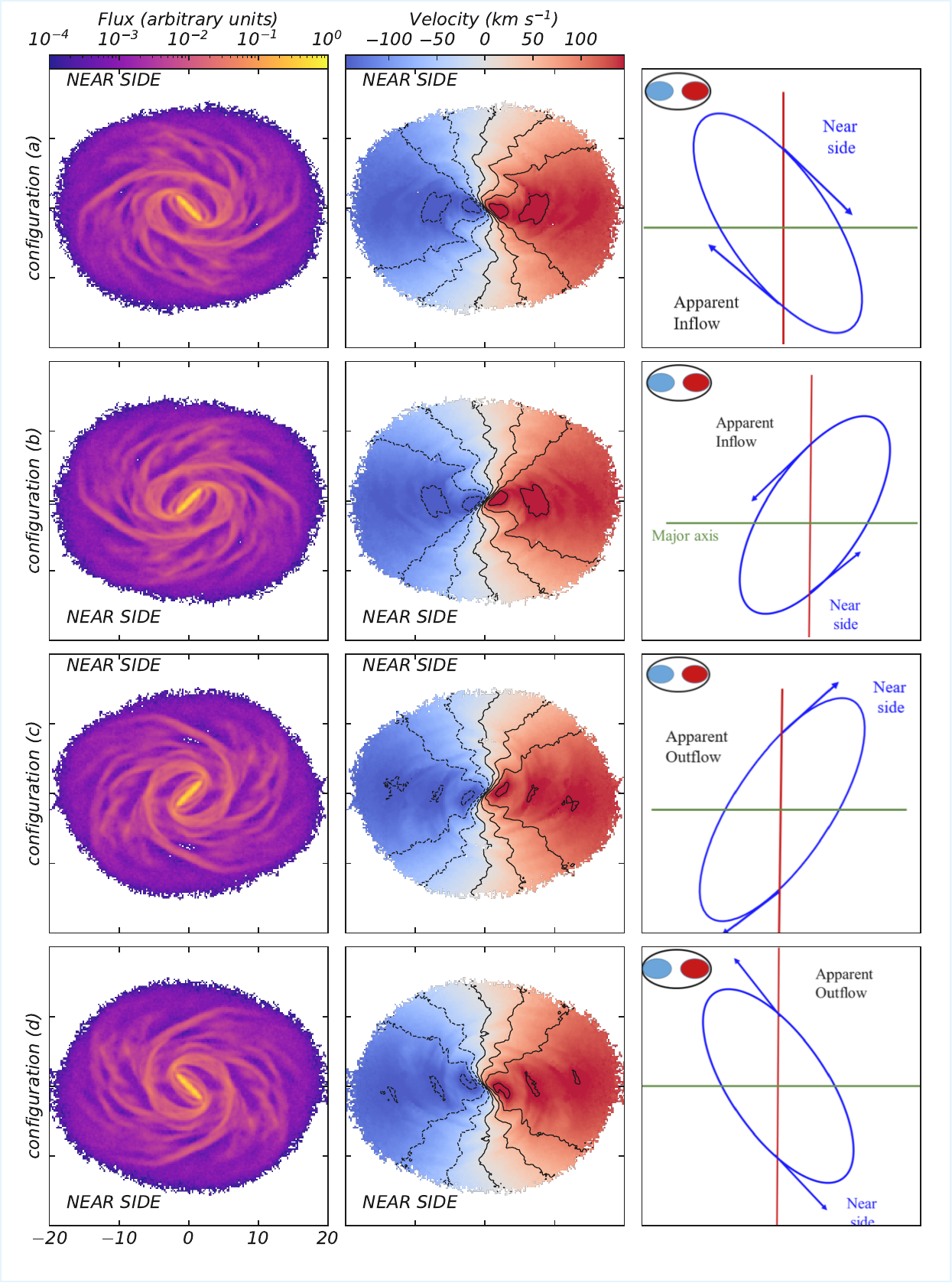}
        \caption{Each row corresponds to one configuration. Left column: moment 0 maps. Middle column: moment 1 maps. The gas component has been inclined by $i_{\rm{true}} = 40^\circ$ and rotated so the bar could form a 45$^\circ$ angle with respect to the galaxy kinematic axes. The image parameters and the isovelocity contours are the same as in Fig.~\ref{fig:bar_symaxes_moms}. Right column: the corresponding bar diagram with the radial motion interpretation according to the gas velocity vector projected on the minor axis in red and the major axis is in green.}
    \label{fig:configs}
\end{figure*}

We also studied four configurations shown in the different rows of Fig.~\ref{fig:configs}. The moment 0 and moment 1 maps of the gas are shown in the left and middle columns of Fig.~\ref{fig:configs} for an intermediately inclined disk, $i_{\rm{true}} = 40^\circ$ in different configurations. The gas particles have been rotated so the bar can be oriented at $\pm45^\circ$ from the major and minor kinematic axes once the disk has been inclined. In the moment 1 maps (Fig.~\ref{fig:configs}, middle column), the disk is dominated by rotation but the kinematic major and minor axes are not perpendicular, especially in the bar region. The isovelocity contours (black lines on the moment 1 maps) draw the characteristic $\mathcal{S}$-shaped pattern revealing an oval distortion of the potential along the bar \citep{1995gaco.book.....C}.

The only difference between configurations (a) and (c), and between (b) and (d), is the orientation of the bar. This orientation modifies the projection of the gas velocity vector onto the minor axis, thereby influencing whether the radial velocity is interpreted as inflow or outflow (see right column of Fig.~\ref{fig:configs}). Conversely, configurations (a) and (d), as well as (b) and (c), differ only by which side of the disk is closest to the observer. This distinction affects the interpretation of the radial velocity as it reverses the apparent sense of gas rotation.

We model the gas kinematics in the configurations presented in Fig.~\ref{fig:configs} at inclination angles $i_{\rm{true}}\in[30^\circ,80^\circ]$ with the following input parameters combinations in \texttt{3D-Barolo}:
\begin{itemize}
    \item a free PA and a free inclination angle (Sect~\ref{subsec:freei_freePA}),
    \item a fixed PA and a free inclination angle (Sect~\ref{subsec:freei_fixedPA}),
    \item a free PA and a fixed inclination angle (Sect~\ref{subsec:fixedi_freePA}),
    \item a fixed PA a fixed inclination angle (Sect.~\ref{subsec:fixedi_fixedPA}).
\end{itemize}

\subsection{Results with a free PA and a free inclination angle}\label{subsec:freei_freePA}

\begin{figure*}[ht]
    \centering
     \includegraphics[width=0.94\textwidth]{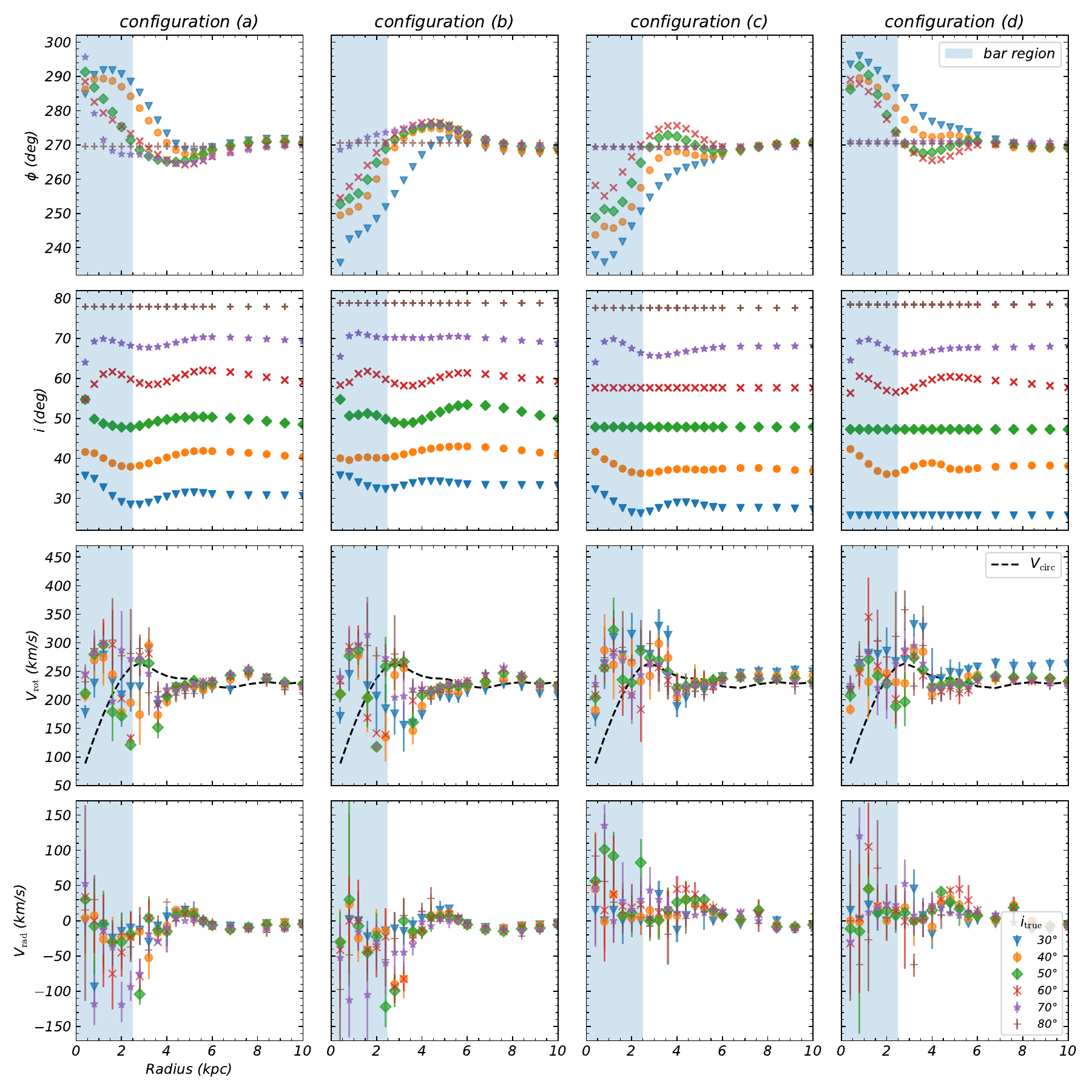}
    \caption{Same as Fig.~\ref{fig:bar_symaxes_params} but each column corresponds to one configuration presented in the rows of Fig.~\ref{fig:configs}.}
    \label{fig:fits_PA_free}
\end{figure*}

The parameters fitted by \texttt{3D-Barolo} with a free PA and a free inclination angle in the different configurations of Fig.~\ref{fig:configs} are plotted in Fig.~\ref{fig:fits_PA_free}.

Depending on the configuration and the true inclination angle, the program uses either a median value or a Bézier interpolation for the geometrical parameters. Bézier interpolation is predominant among the fits of the geometrical parameters, meaning that the dispersion is usually more than $3^\circ$ for the PA and $i$.

The lower the inclination, the further the PA estimated by \texttt{3D-Barolo} in the region $R<4$~kpc is from the expected PA of $\phi=270^\circ$ (first row of Fig.~\ref{fig:fits_PA_free}). This region corresponds to an area larger than the bar region ($R<2.5$~kpc), but errors are significant within this area. Beyond 4~kpc, the computed PA corresponds to the correct value we used as a fixed input in Sect.~\ref{subsec:freei_fixedPA}, $\phi=270^\circ$. However, inside the bar, \texttt{3D-Barolo} tends to oppose the PA to the bar orientation when it is at $\pm45^\circ$. This is likely due to the influence of the spiral structure around the bar. For a highly inclined disk ($i_{\rm true}=80^\circ$ in all configurations and also $i_{\rm true}=70^\circ$ in configurations (c) and (d)), \texttt{3D-Barolo} computes the right PA even in the bar region. The bar is less observable in almost edge-on disks and the algorithm seems to ignore its effect on the geometrical parameters in this case.

When the bar is at $\pm 45^\circ$ from the kinematic axes (first row of Fig.~\ref{fig:fits_PA_free}), the PA varies more than when the bar is along the minor axis but less than when the bar is along the major axis (first row of Fig.~\ref{fig:fits_PA_free}). With a free inclination angle, the maximum PA variation is:
\begin{itemize}
    \item $\Delta\phi_{\rm max}\simeq55^\circ$ for the bar aligned with the major axis,
    \item $\Delta\phi_{\rm max}\simeq36^\circ$ for the bar at $\pm45^\circ$,
    \item $\Delta\phi_{\rm max}\simeq17^\circ$ for the bar aligned with the minor axis.
\end{itemize}
The more the bar affects the tangential velocity, the more \texttt{3D-Barolo} attempts to compensate for this effect by modifying the PA.

Across all four configurations and true inclination angles, the Bézier curves computed for $i$ (second row of Fig.~\ref{fig:fits_PA_free}) follow a similar trend: the inclination angle decreases around the bar end (from $R\sim2$ to 2.5~kpc), and then it oscillates around a value close to $i_{\rm{true}}$. When \texttt{3D-Barolo} finds a median value for the inclination, it is systematically underestimated by a few degrees (between 1.1$^\circ$ and 4.3$^\circ$).

All the rotational curves (third row of Fig.~\ref{fig:fits_PA_free}) should be identical as they represent the deprojected tangential velocity. However, we notice large discrepancies in the bar region, especially for the lowest true inclination angles. This is caused by the wrong estimation of the PA and $i$ in the same region. The bar is about 5~kpc long so we expect the rotation curve to be wrongly estimated for $R<2.5$~kpc. The geometrical parameters have a great influence on the kinematic parameters modeling e.g. in configuration (d) at the lowest inclination $i_{\rm true}=30^\circ$, the rotational curve is wrong even outside the bar region because the inclination angle, found by \texttt{3D-Barolo} during a previous step, was underestimated. They are responsible for the differences between rotation curves in a same configuration or at the same true inclination. The rotational velocity computed by \texttt{3D-Barolo} is similar to the circular velocity $V_{\rm{circ}}$ for $R>4$~kpc (third row of Fig.~\ref{fig:fits_PA_free}), so outside the bar region the gas orbits are circular. However, between 6 and 9~kpc which seems to be in the spiral arm region, \texttt{3D-Barolo} finds a tangential velocity 10-20~\kms higher than the expected circular velocity in configurations (a) and (b).

The opposite signs of the recovered radial velocities (fourth row of Fig.~\ref{fig:fits_PA_free}) reflect the symmetry between the configurations. In the bar region ($R<2.5$~kpc), \texttt{3D-Barolo} retrieves inflow in configurations (a) and (b), whereas their symmetric counterparts, configurations (c) and (d), show outflow, although with smaller absolute radial velocities. These differences arise from the variations in the tangential component induced by the incorrect estimation of the geometrical parameters.

\begin{figure}[ht]
    \centering
    \resizebox{\hsize}{!}{
    \includegraphics[width=0.44\textwidth]{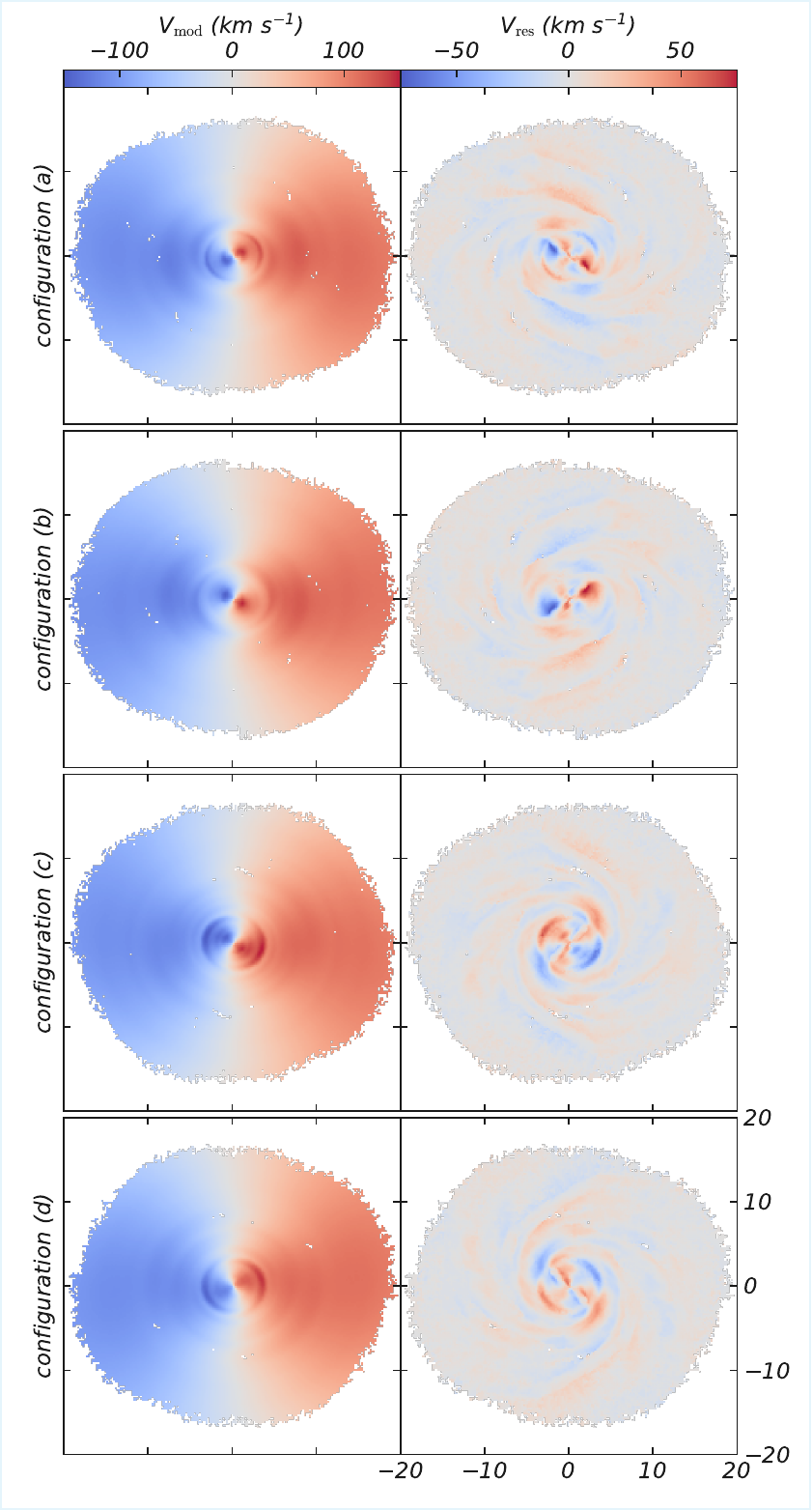}}
    \caption{Rows correspond to the configurations presented in Fig.~\ref{fig:configs} with a disk inclination of $i_{\rm true}=30^\circ$. Left column: modeled moment 1 maps with the corresponding parameters plotted in Fig.~\ref{fig:fits_PA_free} with a free PA and a free inclination angle. Right column: residuals maps ($V_{\rm LoS}-V_{\rm mod}$).}
    \label{fig:mom1_comp}
\end{figure}

The modeled moment 1 maps by \texttt{3D-Barolo} and the residuals maps (model subtracted from the data) at the lowest inclination ($i_{\rm true}=30^\circ$) are presented in Fig.~\ref{fig:mom1_comp}, with each row corresponding to a configuration of Fig.~\ref{fig:configs}. The corresponding parameters are plotted in Fig.~\ref{fig:fits_PA_free} with blue triangles.
We see that the modeled velocity fields do not match the data, especially in the bar region where the residuals are actually bar-shaped, more clearly in configurations (a) and (b) (first two rows of Fig.~\ref{fig:mom1_comp}). The bar-shaped residuals demonstrate that the bar is the primary source of the modeling errors, and that the radial velocities recovered by \texttt{3D-Barolo} do not incorporate the bar-induced elliptical orbits.

\subsection{Results with a fixed PA and a free inclination angle}\label{subsec:freei_fixedPA}

\begin{figure*}[ht]
    \centering
     \includegraphics[width=0.94\textwidth]{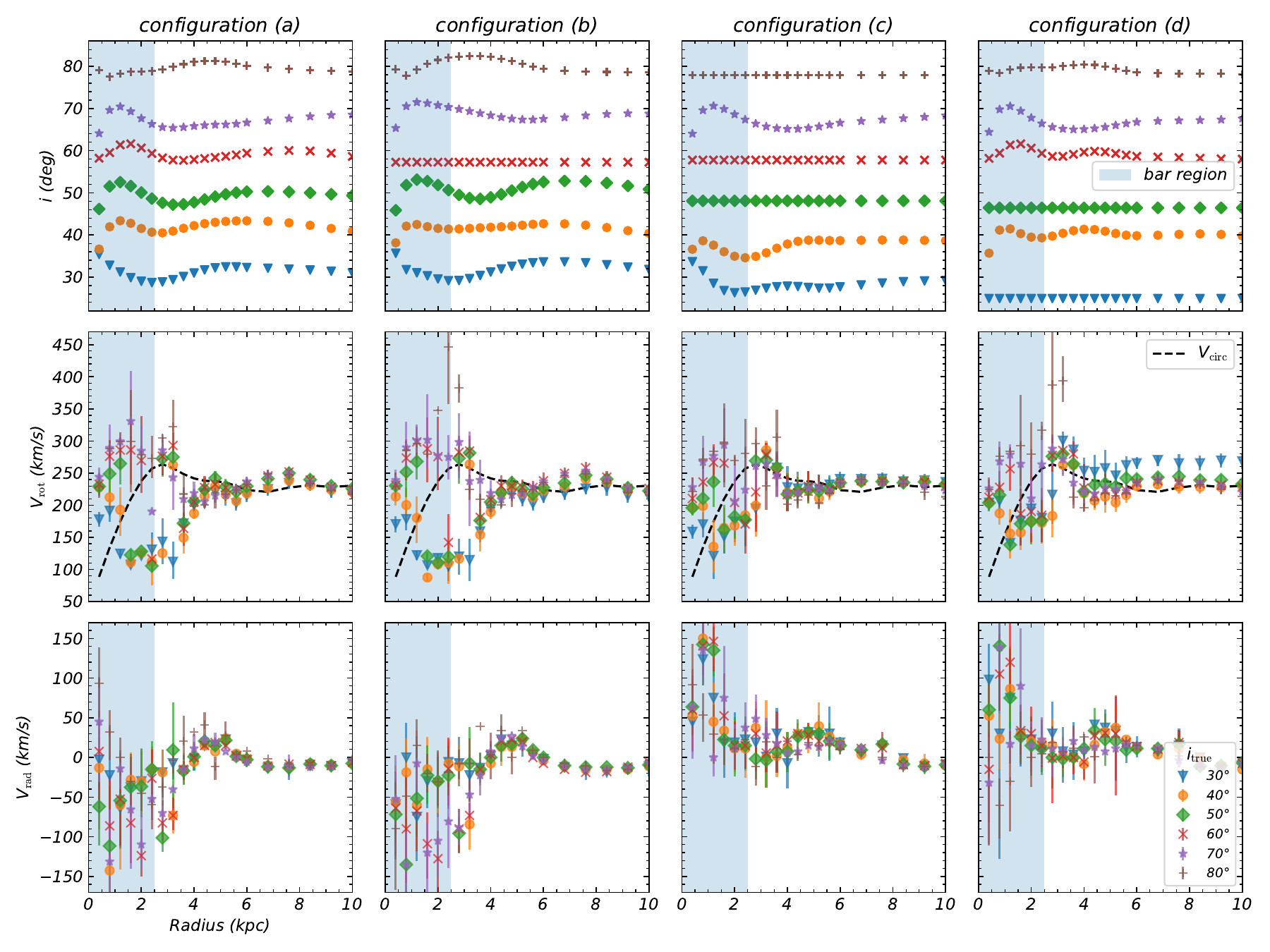}
    \caption{Same as Fig.~\ref{fig:fits_PA_free} but parameters were fitted with a fixed PA of $\phi=270^\circ$ and a free inclination angle.}
    \label{fig:fits_PA_fixed}
\end{figure*}

We fix the PA at $\phi=270^\circ$ to prevent the code from misinterpreting the bar-induced distortions as a kinematic warp, as often occurs when PA is left free (see \citealt{2000AstL...26..565M} and Sect.~\ref{subsec:freei_freePA}). The parameters fitted by \texttt{3D-Barolo} with a fixed PA and a free inclination, in the configurations presented in Fig.~\ref{fig:configs}, are plotted in Fig.~\ref{fig:fits_PA_fixed}.

For the lowest inclination angles ($30^\circ\leq i_{\rm{true}}\leq 50^\circ$), we observe that $V_{\rm{rot}}$ decreases from the galactic center to around 2.5~kpc, the end of the bar. $V_{\rm{rot}}$ reaches lower values in configurations (a) and (b) (apparent inflow configurations) but the trend is similar for the same inclinations in configurations (c) and (d). On the opposite, for the highest inclinations ($60^\circ\leq i_{\rm{true}}\leq 80^\circ$), the fitted values of $V_{\rm{rot}}$ are higher than the expected circular velocity in the same region for all configurations. Especially at $i_{\rm{true}}=80^\circ$, in the bar region, the rotational velocity is higher than 300~\kms in configurations (b) and (d). In configuration (d), for the lowest inclined disk ($i_{\rm{true}}=30^\circ$), \texttt{3D-Barolo} computes higher values of $V_{\rm{rot}}$ due to its wrong estimation of the inclination angle in this case. This error was not caused by the PA estimation as this case is similar to the results obtained with a free PA (see third row, last column panel of Fig.~\ref{fig:fits_PA_free}).

In configurations (a) and (b), \texttt{3D-Barolo} finds an inflow within $R<4$~kpc for all inclination angles, consistent with the bar location ($R<2.5$~kpc) and projection of the gas velocity vector along the minor axis (right column of Fig.~\ref{fig:configs}). In configurations (c) and (d), \texttt{3D-Barolo} detects an outflow within the same radius for all inclination angles $i_{\rm{true}}\leq 70^\circ$ which also matches the expected projection effects. The unexpected sign of radial velocities for $i_{\rm{true}}=80^\circ$ at $R<4$~kpc are due to the wrong estimation of the rotation curve at this inclination. The radial velocities computed by \texttt{3D-Barolo} reach values exceeding 50\% of $V_{\rm{rot}}$, indicating that projection effects can strongly bias the interpretation of radial motions. The apparent inflows and outflows correspond to high velocities of up to $|V_{\rm{rad}}| \sim 150$~\kms in the central regions. The bar also induces a great radial component closer to its extremities. Beyond 4~kpc, the radial velocity is low and similar for all configurations, confirming that the bar is responsible for the observed variations within 4~kpc. Although there is no bar beyond 2.5~kpc, we observe that the error propagates beyond the bar region.

\subsection{Results with a free PA and a fixed inclination angle}\label{subsec:fixedi_freePA}

In this section, we force \texttt{3D-Barolo} to use the true inclination, fixed for all rings, instead of fitting it. The PA, however, is left free and may vary with radius. This allows us to test whether knowing the correct inclination alone is sufficient for accurate kinematic modeling, or whether the free PA still leads to a misinterpretation of bar-driven motions. Parameters fitted by \texttt{3D-Barolo} in the configurations of Fig.~\ref{fig:configs} with a fixed inclination angle are plotted in Fig.~\ref{fig:fits_inc_fixed}.

The computed PA values (first row of Fig.~\ref{fig:fits_inc_fixed}) with a fixed inclination angle are similar to the ones obtained with a free $i$ (first row of Fig.~\ref{fig:fits_PA_free}).

Fixing the inclination produces a stronger overestimation of the tangential velocity at high inclination in configurations (c) and (d) (second row of Fig.~\ref{fig:fits_inc_fixed}). Once the inclination can no longer adjust, the fit compensates for the bar-induced non-circular motions by artificially increasing $V_{\rm rot}$. In the other cases, the rotational curves are similar than the ones obtained with a free $i$ (third row of Fig.~\ref{fig:fits_PA_free}).

In the case with a fixed inclination angle and a free PA, the recovered radial velocities (third row of Fig.~\ref{fig:fits_inc_fixed}) are similar to those obtained with a free inclination angle (fourth row of Fig.~\ref{fig:fits_PA_free}) in most configurations. However, for configurations (c) and (d) at $i_{\rm true}=70^\circ$, where \texttt{3D-Barolo} previously converged to a constant median PA of $\phi=270^\circ$ when both PA and $i$ were free (see Sect.~\ref{subsec:freei_freePA}), fixing the inclination now forces the code to fit the PA with a Bézier curve, which constitutes an incorrect estimation.

\subsection{Results with a fixed PA and a fixed inclination angle}\label{subsec:fixedi_fixedPA}

\begin{figure*}[ht]
    \centering
     \includegraphics[width=0.94\textwidth]{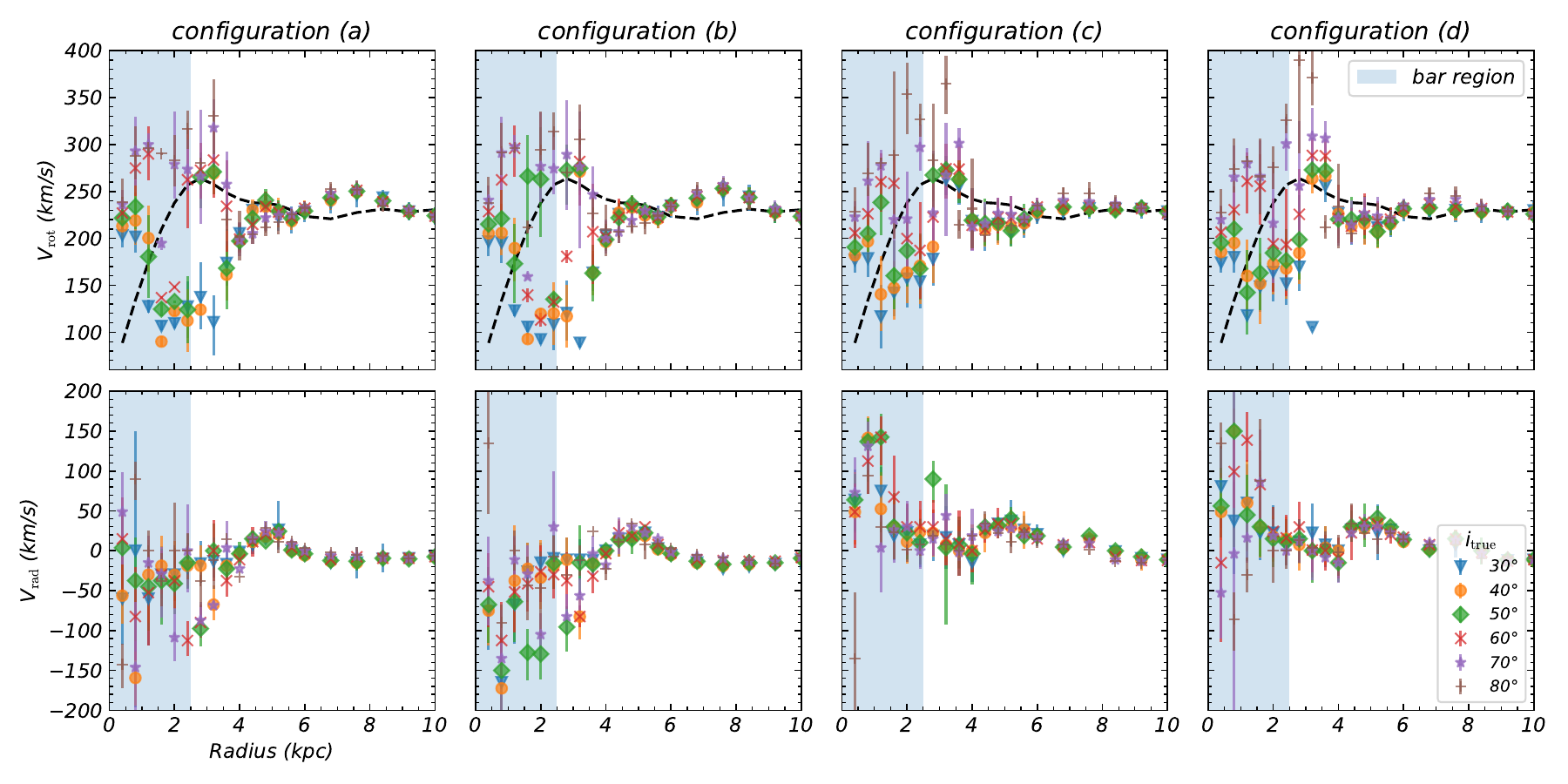}
    \caption{Same as Fig.~\ref{fig:fits_PA_free} but parameters were fitted with a fixed PA of $\phi = 270^\circ$ and a fixed inclination angle.}
    \label{fig:fits_INC_PA_fixed}
\end{figure*}

In general, the geometric parameters ($i$, PA) can be determined assuming that the outer regions of the galaxy disks are circular, unperturbed. Therefore, it is interesting to only fit the kinematic parameters with a fixed PA ($\phi=270^\circ$) and inclination angle ($i=i_{\rm true}$). The rotation curves and the radial velocities derived by \texttt{3D-Barolo} with this combination are plotted on Fig.~\ref{fig:fits_INC_PA_fixed}.

The results are very similar to those obtained with a fixed PA and free inclination angle (see Fig.~\ref{fig:fits_PA_fixed}), indicating that the PA is the most critical geometrical parameter in the kinematics modeling. We note less discrepancy in the tangential velocities outside the bar region with the varying inclination angle when both geometrical parameters are fixed. One difference from the previous input parameters combinations results is that in configurations (a) and (d) -- whose bar's orientation is the same -- two regimes are observed in the rotation curves (first row of Fig.~\ref{fig:fits_INC_PA_fixed}) depending on the inclination angle. For an inclination of $30^\circ\leq i_{\rm true}\leq 50^\circ$, there is a dip feature in the bar region, discussed in Sect.~\ref{subsec:comp}, whereas for $60^\circ\leq i_{\rm true}\leq 80^\circ$ it is either non-existent either less pronounced. Similar features appear in the other configurations but with some unexpected values. Even when they are fixed to their true values, we observe a wrong estimation of the rotation curves, particularly within the bar region.

\section{Discussion}\label{sec:discussion}

In Sect.~\ref{subsec:comb}, we compare the results obtained with different input parameters combinations and evaluate how the bar orientation and projection influence the recovered velocities. We discuss our results with other studies done on observations in Sect.~\ref{subsec:comp}.

\subsection{Comparison of the kinematics modeling at $\pm45^\circ$ bar orientation: influence of input parameters}\label{subsec:comb}

The first limitation of our study concerns the algorithm itself. As noted by \cite{2021ApJ...923..220D}, \texttt{3D-Barolo} assumes an axisymmetric disk, which is not an adequate description for a barred spiral galaxy. When used on observations, \texttt{3D-Barolo} usually fits the geometrical parameters of the galactic disk but even in the best cases, where the PA and/or the inclination angle are known and fixed to a constant value, one cannot escape the bar's effects.

The lower the disk inclination, the more \texttt{3D-Barolo} changes the PA between the galactic center and the external parts of the disk. This phenomenon of compensation for the complexity in the gas kinematics due to the presence of the bar is observed whether the inclination angle is fixed (first row of Fig.~\ref{fig:fits_inc_fixed}) or not (first row of Fig.~\ref{fig:fits_PA_free}).

In most cases, when a Bézier curve fits the inclination angle, the mean value computed by \texttt{3D-Barolo} outside the bar region is closer to the true value when the PA is fixed at $\phi=270^\circ$ (first row of Fig.~\ref{fig:fits_PA_fixed}) than when it is left free (second row of Fig.~\ref{fig:fits_PA_free}). However, there are some exceptions in configurations (c) and (d). \texttt{3D-Barolo} tends to underestimate the median inclination more compared to the values obtained when the PA is free to vary (between 1.9$^\circ$ and 5.1$^\circ$ versus 1.1$^\circ$ and 4.3$^\circ$).

The rotation curves are similar whether the PA is left free (third row of Fig.~\ref{fig:fits_PA_free}) or fixed (the second row of Fig.~\ref{fig:fits_PA_fixed}) outside $R>4$~kpc. At $i_{\rm true}=80^\circ$, there are unexpected high values of $V_{\rm rot}$ in configurations (b) and (d) (second row of Fig.~\ref{fig:fits_PA_fixed}) around $R\sim 3$~kpc that are not there when the PA is free (third row of Fig.~\ref{fig:fits_PA_free}) as the computed inclination was a constant median value.

The apparent radial velocities obtained by \texttt{3D-Barolo} in the bar region when the PA is left free (fourth row of Fig.~\ref{fig:fits_PA_free}), are lower than with a fixed PA of $\phi=270^\circ$ (third row of Fig.~\ref{fig:fits_PA_fixed}).

\begin{figure}[ht]
    \centering
    \resizebox{\hsize}{!}{
    \includegraphics[width=0.44\textwidth]{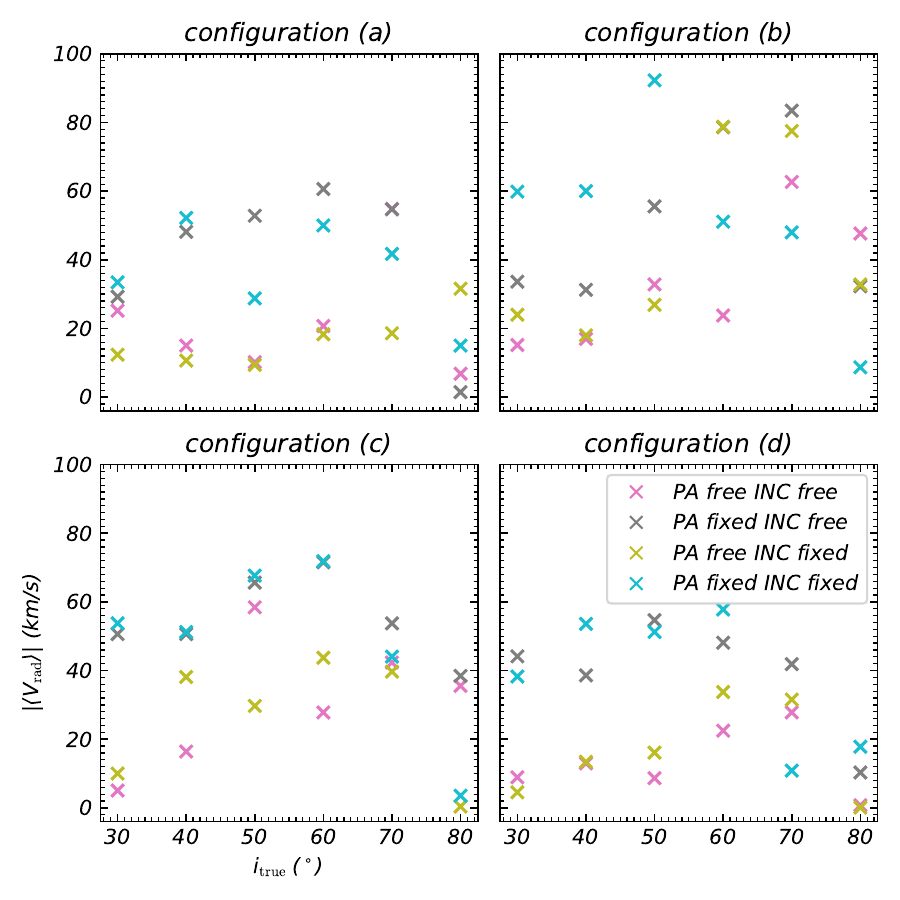}}
    \caption{Absolute mean radial velocity within $R<2.5$~kpc as a function of the true inclination for all four configurations of Fig.~\ref{fig:configs} with the bar at $\pm 45^\circ$ from the kinematic axes. Each color corresponds to one combination of input parameters.}
    \label{fig:mean_vrad}
\end{figure}

The mean value of $V_{\rm{rad}}$ within the bar region as a function of the true inclination angle for each configuration is shown in Fig.~\ref{fig:mean_vrad}.

In general, apparent radial velocities increase with inclination for $i_{\rm{true}}\leq60^\circ$. At low inclination angles, the more inclined the galactic disk is, the more overestimated radial flows are. 

For highly inclined disk, the apparent radial velocities decrease, reaching values close to 0 for $i_{\rm{true}}=80^\circ$ in certain cases. The values change a lot with the configurations and the input parameters combination. This confirms that the bar affects the recovery of both geometrical and kinematic parameters, introducing substantial uncertainty in the apparent radial velocities inferred by \texttt{3D-Barolo}. The low radial velocities force the model to absorb the bar-driven non-circular motions into the tangential component, producing the artificially high values of $V_{\rm{rot}}$ at this inclination (second rows of Fig.~\ref{fig:fits_PA_fixed}, Fig.~\ref{fig:fits_inc_fixed} and first row of Fig.~\ref{fig:fits_INC_PA_fixed}). At this high inclination, \texttt{3D-Barolo} attempts to fit circular orbits, effectively ignoring the non-circular motions produced by the bar. 

We note that the highest absolute radial velocities are retrieved when the PA is fixed and the inclination angle is left free (gray markers in Fig.~\ref{fig:fits_PA_fixed}). In this case, the mean radial velocities are significant as they range from $|\langle V_{\rm rad}\rangle|\simeq$~29 to 84~\kms for $i_{\rm{true}}\leq70^\circ$. Since the spurious mass flow rates scale with velocity, this could result in substantial overestimates of the spurious mass flow rates. Because the bar breaks axisymmetry, models that impose a fixed PA or assume circular motions misinterpret the bar-driven velocity field, often yielding spurious detections of radial flows.

\subsection{Comparison with previous studies}\label{subsec:comp}

\cite{2004ApJ...605..183W} already demonstrated that the non-axisymmetric nature of spiral galaxies can lead to false detections of inflows in the inner disk. Using CO and HI interferometric data, they found that elliptical streaming within the bar dominates non-circular motions, leaving no significant inflow. Unlike 2D Fourier decomposition that was used by \cite{2004ApJ...605..183W}, 3D-Barolo fits the full spectral cube. In this work, we therefore examined how such 3D modeling responds to the non-axisymmetric motions generated by a bar, and how it still leads to spurious inflow signatures.

In a large HI survey, \cite{2021ApJ...923..220D} found that the strongest apparent radial velocities were always associated with bars or spiral arms, i.e. non-axisymmetric structures. Our results confirm that bars systematically increase the uncertainty in $V_{\mathrm{rad}}$ and can easily mimic radial flows.

Even when the geometrical parameters were set to their true values, the rotation curve estimations were incorrect due to the bar. Similar results were obtained by \cite{2025ApJ...980..146L} using a tilted-ring method without a radial velocity while fixing the inclination at $i=45^\circ$ and the PA at $\phi=0^\circ$. They found that the rotation curves were strictly dependent on the bar's orientation in the sky plane. Their simulations at $\Delta\phi=\pm45^\circ$ (see their Fig.~1) respectively match our configurations (c) and (a) of Fig.~\ref{fig:configs} because the gas rotates clockwise, with $\Delta\phi$ the difference between the galaxy's PA and the bar's. In these cases, they identified a 'dip' feature in the rotation curves within the bar region: a central value higher than the circular velocity followed by a local minimum still within the bar region. We observed the same feature on the rotation curves derived for $30^\circ\leq i_{\rm true}\leq 50^\circ$ (Fig.~\ref{fig:fits_INC_PA_fixed}) in all configurations, unlike \cite{2025ApJ...980..146L} who found a peak when $\Delta \phi$ is negative that we do not retrieve. They also studied the rotation curves of 45 galaxies from the PHANGS-ALMA survey, including 29 barred galaxies. The dip feature was observed in 9 of the 11 galaxies with $|\Delta\phi\leq40^\circ|$. According to \cite{2025ApJ...980..146L}, the strength of the bar is related to the strength of the dip. Although the feature is not dominant in the sample of barred galaxies, its absence in the sample of non-barred galaxies shows that it is indeed bar-induced. They suppose this is due to bar orientation. Additionally, identifying such features is complicated by other sources of changes in rotation curves, such as spiral arms or massive bulges within the bar region. Besides that, they proposed a misaligned ellipses model showing how projection creates velocity artifacts that can be misinterpreted as radial flows, as we demonstrated. Because this geometric model alone cannot fully recover the intrinsic circular velocity, they further introduced an empirical, first-order correction to the rotation curve. This correction is based on calibrating the systematic deviations induced by the bar as a function of bar-disk position angle offset and radius.

In this work, we used a simulated galaxy but several observed galaxies that have been modeled with \texttt{3D-Barolo} show the pattern we highlight in this study: apparent inflows or outflows in barred regions are often projection artifacts.

NGC~1808, a strongly barred late-type spiral galaxy \citep{1991rc3..book.....D}, is known for having large-scale outflows perpendicular to the disk, due to star formation \citep{2016ApJ...823...68S}. 
\cite{2021A&A...656A..60A} derived CO kinematics with \texttt{3D-Barolo} and found large radial velocities in the central 500 pc, where a nuclear spiral is present. From \texttt{3D-Barolo}'s interpretation, there was an apparent inflow. An inflow was confirmed by gravitational torque analysis, which showed negative azimuthally-averaged and gas-weighted torques consistent with true inflow, but with lower amplitude. At smaller scales, they identify the variations of the PA as perturbations caused by the decoupling of the torus. In our case, the scale is too large to be affected by nuclear structures like this, but this result shows that a changing PA could be attributed to many physical phenomena as a warped disk, a bar that has not been identified or a torus at smaller scales. A similar case is the barred galaxy NGC~613, in which \cite{2019A&A...632A..33A} observed and characterized both a radial inflow through gravity torques, and an outflow perpendicular to the disk.

In contrast, in NGC~7172, \cite{2023A&A...675A..88A} reported an apparent radial molecular outflow in the bar region, whereas gravitational torque analysis revealed an inflow instead. Using ALMA CO(3–2) data, they found significant non-circular motions within the molecular ring, associated with an $\mathcal{S}$-shaped velocity pattern consistent with an inner Lindblad resonance, as well as an ionized gas outflow, perpendicular to the disk. When radial motion was not included in the \texttt{3D-Barolo} model, non-axisymmetric residuals appeared—precisely where the molecular ring is observed—while the modeled rotation curve deviated from the data both in the inner disk and at the edges of the ring ($R\sim3$–$6$~kpc). Adding a radial component improved the fit though spurious velocities ($V_{\mathrm{rad}}\sim50$~\kms) persisted well beyond the ring. The authors argued that a central bar could explain these kinematic signatures: it produces an apparent molecular outflow that is in fact a projection artifact, while the true motion—confirmed by simulations and torque analysis—is a molecular inflow driven by the bar’s gravitational forces. Thus, as in our results, the bar-induced radial velocities propagate beyond the bar region.

\cite{2024A&A...686A..46E} found similar effects: adding a radial component improved the fit near the nucleus, yet \texttt{3D-Barolo} still misinterpreted velocities in the bar region. Because of the high disk inclination, the presence of a bar remains uncertain, and the authors considered the measured radial velocities as upper limits.

Finally, \cite{2014A&A...567A.125G,2019A&A...632A..61G} used ALMA data to model the CO kinematics in the circumnuclear disk of the barred galaxy NGC 1068. Fourier decomposition revealed inflow along the stellar bar. They also detected a molecular outflow, consistent with AGN feeding and feedback. However, the apparent inflow outside the bar was attributed to projection effects combining bar and spiral arms.

Across these different systems, a consistent picture emerges: radial velocities derived from axisymmetric models in barred regions should be considered upper limits rather than direct measurements of gas inflow or outflow. Combining kinematics with gravitational torque analysis or multi-tracer observations is essential to distinguish true radial flows from projection-induced artifacts.

\begin{figure}[ht]
    \centering
    \resizebox{\hsize}{!}{
    \includegraphics[width=0.44\textwidth]{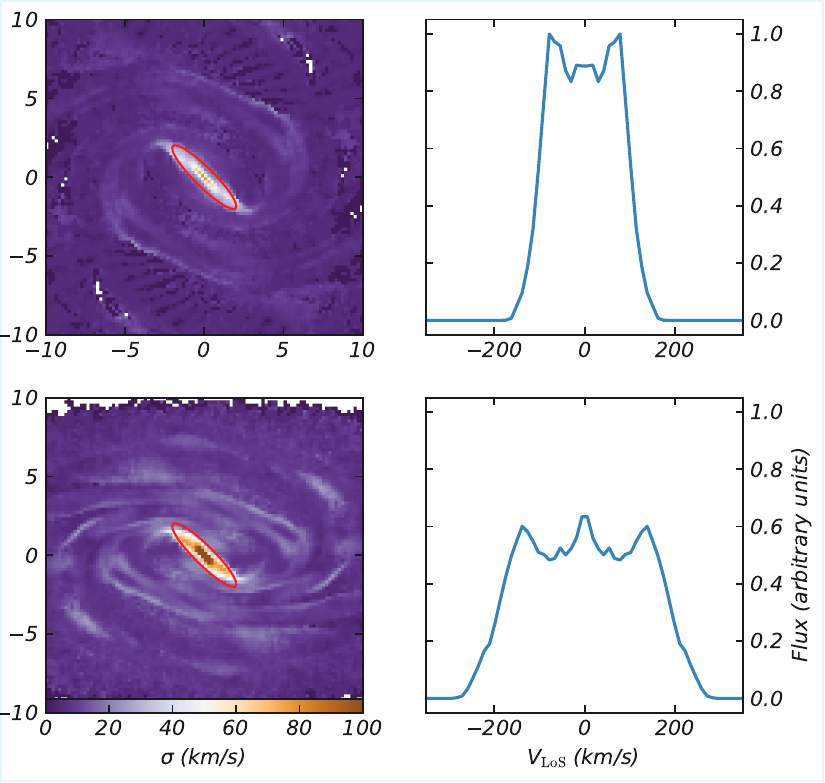}}
    \caption{Left column: moment 2 maps (velocity dispersion). Right column: averaged $z$-profiles within the red ellipse drawn on the moment 2 maps. Top row: gSb model with an inclination of $i_{\rm true}=30^\circ$. Bottom row: same model with an inclination of $i_{\rm true}=60^\circ$, both shown for configuration (a) (see Fig.~\ref{fig:configs}).}
    \label{fig:Zprofiles}
\end{figure}

\begin{figure}[ht]
    \centering
    \resizebox{\hsize}{!}{
    \includegraphics[width=0.44\textwidth]{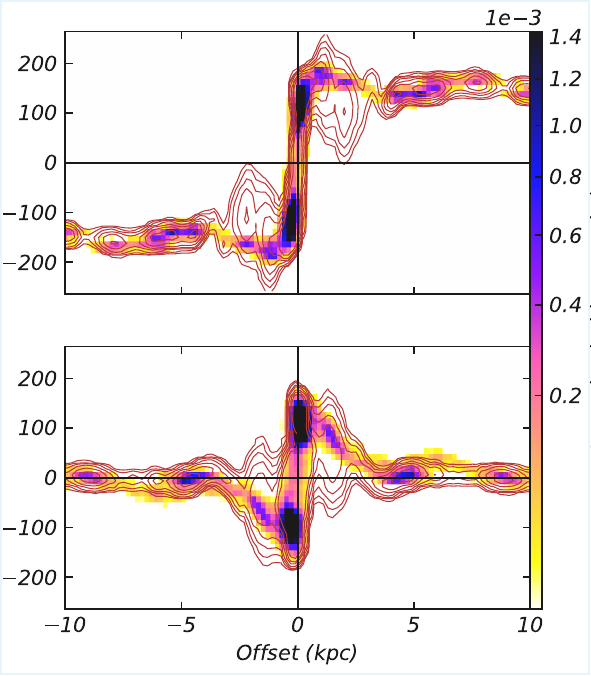}}
    \caption{Position-velocity diagrams along the major axis at $\phi=270^\circ$ (top panel) and minor axis at $\phi=0^\circ$ (bottom panel) for gSb inclined at $i_{\rm true}=40^\circ$ in configuration (a) of Fig.~\ref{fig:configs}. The red contours are the best fit from \texttt{3D-Barolo} obtained with a free PA and a free inclination angle (see Sect.~\ref{subsec:freei_freePA}).}
    \label{fig:PV}
\end{figure}

\cite{2026A&A...705A..51D} modeled the CO kinematics in three Seyfert galaxies: NGC~6860 (an (R')SB(r)b galaxy), Mrk~915 (Sa), and MCG-01-24-012 (SAB(rs)c), using ALMA data. They identified regions with double-peak (DP) spectra, which arise when the LoS intersects multiple gas components with distinct kinematics. \cite{2023A&A...670A..46M} found that bars alone can generate a DP signature for certain viewing angles, without inflows, outflows, or mergers. Their study also showed that DP signatures can simply result from the central shape of the rotation curve or the presence of multiple gas components. Thus, residuals from a rotating-disk model alone do not provide sufficient information to distinguish whether DP profiles originate from separate gas components along the LoS or from velocity gradients induced by non-axisymmetric motions such as those from the bar. In the gSb model, we also find double and triple-peak profiles within the bar, despite the absence of any imposed inflow or outflow (Fig.~\ref{fig:Zprofiles}). Their occurrence is strongly inclination-dependent and arises purely from the projection of elliptical bar orbits. This demonstrates that multiple spectral components cannot be uniquely attributed to radial flows without additional constraints.
\cite{2026A&A...705A..51D} additionally reported non-circular motions in position-velocity (PV) diagrams along the minor axis (see their Fig.~6), which they attributed to disturbed CO kinematics. In our simulated barred galaxy, similar features arise naturally from the projection of bar-driven elliptical orbits (Fig.~\ref{fig:PV}), without any actual radial inflow or outflow. This suggests that some of the observed signatures may also be consistent with non-axisymmetric streaming motions. Consequently, part of the radial flows reported in the three Seyfert galaxies, could arise from this degeneracy between true radial motions and the LoS projection of elliptical bar orbits.

We showed that projection effects can induce a misinterpretation of radial motions in barred spiral galaxies. Identifying non-axisymmetric structures is therefore a prerequisite before interpreting apparent radial motions as genuine inflows or outflows. In particular, true radial inflows of gas are expected to be weak on average. They can be reliably measured through gravitational torque computations or by comparing them with numerical simulations that match observations. The latter approach can in principle provide detailed constraints, but in practice it requires constructing dynamical models that closely reproduce the observed morphology and kinematics, which is challenging. In addition, kinematic methods based on gas flows along bar dust lanes, calibrated using numerical simulations, offer another way to estimate inflow rates (e.g. \cite{2019MNRAS.484.1213S, 2023MNRAS.523.2918S}). These methods assume that, in the frame co-rotating with the bar, the gas velocity vectors are approximately aligned with the dust lanes. They further rely on simulations to calibrate the fraction of gas that overshoots the nuclear ring and join the other bar lane. As a result, the dust lanes approach provides valuable constraints on non-circular motions and on the gas angular-momentum loss. However, translating the kinematics measured along dust lanes into net inflow rates still depends on the extent to which the adopted simulations capture the true gas dynamics of individual galaxies.

\section{Conclusions}\label{sec:conclu}

We used a simulated giant barred spiral galaxy from the GalMer database, symmetrized and convolved with a Gaussian beam to produce a mock data cube. Modeling its gas kinematics with the tilted-ring algorithm \texttt{3D-Barolo} demonstrates that the bar induces significant errors in the recovered rotational velocity, which in turn lead to spurious radial velocities and potential misinterpretations of radial flows.

We explored different configurations by varying the bar orientation in the plane of the sky, aligned with the kinematic axes and at $\pm45^\circ$ from them, and by changing the near side of the disk through inclinations $i_{\rm{true}} \in [30^\circ,80^\circ]$ in $10^\circ$ steps. When the bar is not aligned with the major axis, \texttt{3D-Barolo} computes non-zero radial velocities corresponding to the projection of gas motions along elliptical orbits. Apparent inflows or outflows arise depending on the specific configuration due to the projection of the radial component along the minor axis.

We fitted the PA, inclination angle, tangential and radial velocities with \texttt{3D-Barolo}. The algorithm often adjusts the PA in the bar region to compensate for non-circular motions, while the fitted inclination angles generally follow Bézier curves with mean values close to the true input. The rotation curves are accurately recovered outside the bar, but within the bar region, \texttt{3D-Barolo} systematically fails to recover the rotational velocity.

To compensate for this, the algorithm computes significant radial velocities, up to $|V_{\rm{rad}}| \sim 150$~\kms at some radii within the bar region, independently of the bar orientation and the near side, especially when the PA is not fitted. The mean radial velocity within the bar ($R<2.5$~kpc) can reach 84~\kms when the PA is fixed to its correct value ($\phi=270^\circ$). These projection-induced artifacts may lead to spurious mass flow rates and even inversions of the inferred flow direction, highlighting the risk of misinterpreting inflows and outflows when the underlying motions are dominated by elliptical bar orbits. Because our study is based on a single barred galaxy model and a specific kinematic fitting algorithm, the bias amplitudes are only approximate.

Overall, our results demonstrate that bar orientation and projection alone can produce apparent radial motions that mimic inflows or outflows, even in the absence of any true radial gas transport. This highlights the risk of overestimating mass flow rates or misinterpreting the nature of radial flows when relying solely on axisymmetric kinematic models like \texttt{3D-Barolo}. Accurate assessments of gas dynamics in barred galaxies therefore require complementary morphological and dynamical analyses, such as gravitational torque calculations, to distinguish genuine radial motions from projection-induced artifacts.

\begin{acknowledgements}
    This work was supported by the Programme National Cosmology et Galaxies (PNCG) of CNRS/INSU with INP and IN2P3, co- funded by CEA and CNES. The data were processed using the Gildas package. We made use of the NASA/IPAC Extragalactic Database (NED), and of the HyperLeda database (http://leda.univ-lyon1.fr). This paper made use of GalMer data (http://galmer.obspm.fr). This work made use of Astropy (http://www.astropy.org) a community-developed core Python package and an ecosystem of tools and resources for astronomy. We used the algorithm \texttt{3D-Barolo} \citep{2015MNRAS.451.3021D, 2021ApJ...923..220D}.
\end{acknowledgements}

\bibliography{biblio}{}

\begin{thebibliography}{42}
\expandafter\ifx\csname natexlab\endcsname\relax\def\natexlab#1{#1}\fi

\bibitem[{{Alonso Herrero} {et~al.}(2023){Alonso Herrero},
  {Garc{\'\i}a-Burillo}, {Pereira-Santaella}, {Shimizu}, {Combes}, {Hicks},
  {Davies}, {Ramos Almeida}, {Garc{\'\i}a-Bernete}, {H{\"o}nig}, {Levenson},
  {Packham}, {Bellocchi}, {Hunt}, {Imanishi}, {Ricci}, \&
  {Roche}}]{2023A&A...675A..88A}
{Alonso Herrero}, A., {Garc{\'\i}a-Burillo}, S., {Pereira-Santaella}, M.,
  {et~al.} 2023, \aap, 675, A88

\bibitem[{{Athanassoula}(1992)}]{1992MNRAS.259..328A}
{Athanassoula}, E. 1992, \mnras, 259, 328

\bibitem[{{Audibert} {et~al.}(2019){Audibert}, {Combes}, {Garc{\'\i}a-Burillo},
  {Hunt}, {Eckart}, {Aalto}, {Casasola}, {Boone}, {Krips}, {Viti}, \&
  et~al.}]{2019A&A...632A..33A}
{Audibert}, A., {Combes}, F., {Garc{\'\i}a-Burillo}, S., {et~al.} 2019, \aap,
  632, A33

\bibitem[{{Audibert} {et~al.}(2021){Audibert}, {Combes}, {Garc{\'\i}a-Burillo},
  {Hunt}, {Eckart}, {Aalto}, {Casasola}, {Boone}, {Krips}, {Viti}, {Muller},
  {Dasyra}, {van der Werf}, \& {Mart{\'\i}n}}]{2021A&A...656A..60A}
{Audibert}, A., {Combes}, F., {Garc{\'\i}a-Burillo}, S., {et~al.} 2021, \aap,
  656, A60

\bibitem[{{Binney} \& {Tremaine}(1987)}]{1987gady.book.....B}
{Binney}, J. \& {Tremaine}, S. 1987, {Galactic dynamics}

\bibitem[{{Blackman} \& {Pence}(1982)}]{1982MNRAS.198..517B}
{Blackman}, C.~P. \& {Pence}, W.~D. 1982, \mnras, 198, 517

\bibitem[{{Bournaud} \& {Combes}(2002)}]{2002A&A...392...83B}
{Bournaud}, F. \& {Combes}, F. 2002, \aap, 392, 83

\bibitem[{{Buta} \& {Combes}(1996)}]{Buta1996}
{Buta}, R. \& {Combes}, F. 1996, \fcp, 17, 95

\bibitem[{{Casasola} {et~al.}(2008){Casasola}, {Garcia-Burillo}, {Combes},
  {Hunt}, {Krips}, {Schinnerer}, {Baker}, {Boone}, {Eckart}, {Leon}, {Neri},
  {Tacconi}, \& {.}}]{2008arXiv0811.1971C}
{Casasola}, V., {Garcia-Burillo}, S., {Combes}, F., {et~al.} 2008, arXiv
  e-prints, arXiv:0811.1971

\bibitem[{{Casasola} {et~al.}(2011){Casasola}, {Hunt}, {Combes},
  {Garc{\'\i}a-Burillo}, \& {Neri}}]{2011A&A...527A..92C}
{Casasola}, V., {Hunt}, L.~K., {Combes}, F., {Garc{\'\i}a-Burillo}, S., \&
  {Neri}, R. 2011, \aap, 527, A92

\bibitem[{{Chilingarian} {et~al.}(2010){Chilingarian}, {Di Matteo}, {Combes},
  {Melchior}, \& {Semelin}}]{2010A&A...518A..61C}
{Chilingarian}, I.~V., {Di Matteo}, P., {Combes}, F., {Melchior}, A.~L., \&
  {Semelin}, B. 2010, \aap, 518, A61

\bibitem[{{Combes}(2023)}]{2023Galax..11..120C}
{Combes}, F. 2023, Galaxies, 11, 120

\bibitem[{{Combes} {et~al.}(1995){Combes}, {Boisse}, {Mazure}, {Blanchard}, \&
  {Seymour}}]{1995gaco.book.....C}
{Combes}, F., {Boisse}, P., {Mazure}, A., {Blanchard}, A., \& {Seymour}, M.
  1995, {Galaxies and Cosmology}

\bibitem[{{Combes} {et~al.}(2014){Combes}, {Garc{\'\i}a-Burillo}, {Casasola},
  {Hunt}, {Krips}, {Baker}, {Boone}, {Eckart}, {Marquez}, {Neri}, {Schinnerer},
  \& {Tacconi}}]{2014A&A...565A..97C}
{Combes}, F., {Garc{\'\i}a-Burillo}, S., {Casasola}, V., {et~al.} 2014, \aap,
  565, A97

\bibitem[{{Contopoulos} \& {Grosbol}(1989)}]{1989A&ARv...1..261C}
{Contopoulos}, G. \& {Grosbol}, P. 1989, \aapr, 1, 261

\bibitem[{{Contopoulos} \& {Papayannopoulos}(1980)}]{1980A&A....92...33C}
{Contopoulos}, G. \& {Papayannopoulos}, T. 1980, \aap, 92, 33

\bibitem[{{Dall'Agnol de Oliveira} {et~al.}(2026){Dall'Agnol de Oliveira},
  {Storchi-Bergmann}, {Nagar}, {Garcia-Burillo}, {Riffel}, {Wylezalek},
  {Kukreti}, \& {Ramakrishnan}}]{2026A&A...705A..51D}
{Dall'Agnol de Oliveira}, B., {Storchi-Bergmann}, T., {Nagar}, N., {et~al.}
  2026, \aap, 705, A51

\bibitem[{{Davies} {et~al.}(2024){Davies}, {Shimizu}, {Pereira-Santaella},
  {Alonso-Herrero}, {Audibert}, {Bellocchi}, {Boorman}, {Campbell}, {Cao},
  {Combes}, {Delaney}, {D{\'\i}az-Santos}, {Eisenhauer}, {Esparza Arredondo},
  {Feuchtgruber}, {F{\"o}rster Schreiber}, {Fuller}, {Gandhi},
  {Garc{\'\i}a-Bernete}, {Garc{\'\i}a-Burillo}, {Garc{\'\i}a-Lorenzo},
  {Genzel}, {Gillessen}, {Gonz{\'a}lez Mart{\'\i}n}, {Haidar}, {Hermosa
  Mu{\~n}oz}, {Hicks}, {H{\"o}nig}, {Imanishi}, {Izumi}, {Labiano}, {Leist},
  {Levenson}, {Lopez-Rodriguez}, {Lutz}, {Ott}, {Packham}, {Rabien}, {Ramos
  Almeida}, {Ricci}, {Rigopoulou}, {Rosario}, {Rouan}, {Santos}, {Shangguan},
  {Stalevski}, {Sternberg}, {Sturm}, {Tacconi}, {Villar Mart{\'\i}n}, {Ward},
  \& {Zhang}}]{2024A&A...689A.263D}
{Davies}, R., {Shimizu}, T., {Pereira-Santaella}, M., {et~al.} 2024, \aap, 689,
  A263

\bibitem[{{de Vaucouleurs} {et~al.}(1991){de Vaucouleurs}, {de Vaucouleurs},
  {Corwin}, {Buta}, {Paturel}, \& {Fouque}}]{1991rc3..book.....D}
{de Vaucouleurs}, G., {de Vaucouleurs}, A., {Corwin}, Jr., H.~G., {et~al.}
  1991, {Third Reference Catalogue of Bright Galaxies}

\bibitem[{{Di Teodoro} \& {Fraternali}(2015)}]{2015MNRAS.451.3021D}
{Di Teodoro}, E.~M. \& {Fraternali}, F. 2015, \mnras, 451, 3021

\bibitem[{{Di Teodoro} \& {Peek}(2021)}]{2021ApJ...923..220D}
{Di Teodoro}, E.~M. \& {Peek}, J.~E.~G. 2021, \apj, 923, 220

\bibitem[{{Elmegreen} \& {Elmegreen}(1982)}]{1982MNRAS.201.1021E}
{Elmegreen}, D.~M. \& {Elmegreen}, B.~G. 1982, \mnras, 201, 1021

\bibitem[{{Eskridge} \& {Frogel}(1999)}]{1999Ap&SS.269..427E}
{Eskridge}, P.~B. \& {Frogel}, J.~A. 1999, \apss, 269-270, 427

\bibitem[{{Esposito} {et~al.}(2024){Esposito}, {Alonso-Herrero},
  {Garc{\'\i}a-Burillo}, {Casasola}, {Combes}, {Dallacasa}, {Davies},
  {Garc{\'\i}a-Bernete}, {Garc{\'\i}a-Lorenzo}, {Hermosa Mu{\~n}oz}, {de
  Arriba}, {Pereira-Santaella}, {Pozzi}, {Ramos Almeida}, {Shimizu}, {Vallini},
  {Bellocchi}, {Gonz{\'a}lez-Mart{\'\i}n}, {Hicks}, {H{\"o}nig}, {Labiano},
  {Levenson}, {Ricci}, \& {Rosario}}]{2024A&A...686A..46E}
{Esposito}, F., {Alonso-Herrero}, A., {Garc{\'\i}a-Burillo}, S., {et~al.} 2024,
  \aap, 686, A46

\bibitem[{{Garc{\'\i}a-Burillo} {et~al.}(2019){Garc{\'\i}a-Burillo}, {Combes},
  {Ramos Almeida}, {Usero}, {Alonso-Herrero}, {Hunt}, {Rouan}, {Aalto},
  {Querejeta}, {Viti}, {van der Werf}, {Vives-Arias}, {Fuente}, {Colina},
  {Mart{\'\i}n-Pintado}, {Henkel}, {Mart{\'\i}n}, {Krips}, {Gratadour}, {Neri},
  \& {Tacconi}}]{2019A&A...632A..61G}
{Garc{\'\i}a-Burillo}, S., {Combes}, F., {Ramos Almeida}, C., {et~al.} 2019,
  \aap, 632, A61

\bibitem[{{Garc{\'\i}a-Burillo} {et~al.}(2014){Garc{\'\i}a-Burillo}, {Combes},
  {Usero}, {Aalto}, {Krips}, {Viti}, {Alonso-Herrero}, {Hunt}, {Schinnerer},
  {Baker}, {Boone}, {Casasola}, {Colina}, {Costagliola}, {Eckart}, {Fuente},
  {Henkel}, {Labiano}, {Mart{\'\i}n}, {M{\'a}rquez}, {Muller}, {Planesas},
  {Ramos Almeida}, {Spaans}, {Tacconi}, \& {van der
  Werf}}]{2014A&A...567A.125G}
{Garc{\'\i}a-Burillo}, S., {Combes}, F., {Usero}, A., {et~al.} 2014, \aap, 567,
  A125

\bibitem[{{Hermosa Mu{\~n}oz} {et~al.}(2024){Hermosa Mu{\~n}oz},
  {Alonso-Herrero}, {Pereira-Santaella}, {Garc{\'\i}a-Bernete},
  {Garc{\'\i}a-Burillo}, {Garc{\'\i}a-Lorenzo}, {Davies}, {Shimizu},
  {Esparza-Arredondo}, {Hicks}, {Haidar}, {Leist}, {L{\'o}pez-Rodr{\'\i}guez},
  {Ramos Almeida}, {Rosario}, {Zhang}, {Audibert}, {Bellocchi}, {Boorman},
  {Bunker}, {Combes}, {Campbell}, {D{\'\i}az-Santos}, {Fuller}, {Gandhi},
  {Gonz{\'a}lez-Mart{\'\i}n}, {H{\"o}nig}, {Imanishi}, {Izumi}, {Labiano},
  {Levenson}, {Packham}, {Ricci}, {Rigopoulou}, {Rouan}, {Stalevski},
  {Villar-Mart{\'\i}n}, \& {Ward}}]{2024A&A...690A.350H}
{Hermosa Mu{\~n}oz}, L., {Alonso-Herrero}, A., {Pereira-Santaella}, M.,
  {et~al.} 2024, \aap, 690, A350

\bibitem[{{Hopkins} \& {Quataert}(2010)}]{2010MNRAS.407.1529H}
{Hopkins}, P.~F. \& {Quataert}, E. 2010, \mnras, 407, 1529

\bibitem[{{Liu} {et~al.}(2025){Liu}, {Li}, \& {Shen}}]{2025ApJ...980..146L}
{Liu}, J., {Li}, Z., \& {Shen}, J. 2025, \apj, 980, 146

\bibitem[{{Maschmann} {et~al.}(2023){Maschmann}, {Halle}, {Melchior}, {Combes},
  \& {Chilingarian}}]{2023A&A...670A..46M}
{Maschmann}, D., {Halle}, A., {Melchior}, A.-L., {Combes}, F., \&
  {Chilingarian}, I.~V. 2023, \aap, 670, A46

\bibitem[{{Miller} {et~al.}(1970){Miller}, {Prendergast}, \&
  {Quirk}}]{1970ApJ...161..903M}
{Miller}, R.~H., {Prendergast}, K.~H., \& {Quirk}, W.~J. 1970, \apj, 161, 903

\bibitem[{{Moisseev} \& {Mustsevoi}(2000)}]{2000AstL...26..565M}
{Moisseev}, A.~V. \& {Mustsevoi}, V.~V. 2000, Astronomy Letters, 26, 565

\bibitem[{{Pence} \& {Blackman}(1984{\natexlab{a}})}]{1984MNRAS.207....9P}
{Pence}, W.~D. \& {Blackman}, C.~P. 1984{\natexlab{a}}, \mnras, 207, 9

\bibitem[{{Pence} \& {Blackman}(1984{\natexlab{b}})}]{1984MNRAS.210..547P}
{Pence}, W.~D. \& {Blackman}, C.~P. 1984{\natexlab{b}}, \mnras, 210, 547

\bibitem[{{Randriamampandry} {et~al.}(2016){Randriamampandry}, {Deg},
  {Carignan}, {Combes}, \& {Spekkens}}]{2016A&A...594A..86R}
{Randriamampandry}, T.~H., {Deg}, N., {Carignan}, C., {Combes}, F., \&
  {Spekkens}, K. 2016, \aap, 594, A86

\bibitem[{{Salak} {et~al.}(2016){Salak}, {Nakai}, {Hatakeyama}, \&
  {Miyamoto}}]{2016ApJ...823...68S}
{Salak}, D., {Nakai}, N., {Hatakeyama}, T., \& {Miyamoto}, Y. 2016, \apj, 823,
  68

\bibitem[{{Sanders} \& {Huntley}(1976)}]{Sanders1976}
{Sanders}, R.~H. \& {Huntley}, J.~M. 1976, \apj, 209, 53

\bibitem[{{Shlosman} {et~al.}(1989){Shlosman}, {Frank}, \&
  {Begelman}}]{1989Natur.338...45S}
{Shlosman}, I., {Frank}, J., \& {Begelman}, M.~C. 1989, \nat, 338, 45

\bibitem[{{Sormani} \& {Barnes}(2019)}]{2019MNRAS.484.1213S}
{Sormani}, M.~C. \& {Barnes}, A.~T. 2019, \mnras, 484, 1213

\bibitem[{{Sormani} {et~al.}(2023){Sormani}, {Barnes}, {Sun}, {Stuber},
  {Schinnerer}, {Emsellem}, {Leroy}, {Glover}, {Henshaw}, {Meidt}, {Neumann},
  {Querejeta}, {Williams}, {Bigiel}, {Eibensteiner}, {Fragkoudi}, {Levy},
  {Grasha}, {Klessen}, {Kruijssen}, {Neumayer}, {Pinna}, {Rosolowsky}, {Smith},
  {Teng}, {Tress}, \& {Watkins}}]{2023MNRAS.523.2918S}
{Sormani}, M.~C., {Barnes}, A.~T., {Sun}, J., {et~al.} 2023, \mnras, 523, 2918

\bibitem[{{Wong} {et~al.}(2004){Wong}, {Blitz}, \&
  {Bosma}}]{2004ApJ...605..183W}
{Wong}, T., {Blitz}, L., \& {Bosma}, A. 2004, \apj, 605, 183

\bibitem[{{Zhang} {et~al.}(2024){Zhang}, {Packham}, {Hicks}, {Davies},
  {Shimizu}, {Alonso-Herrero}, {Hermosa Mu{\~n}oz}, {Garc{\'\i}a-Bernete},
  {Pereira-Santaella}, {Audibert}, {L{\'o}pez-Rodr{\'\i}guez}, {Bellocchi},
  {Bunker}, {Combes}, {D{\'\i}az-Santos}, {Gandhi}, {Garc{\'\i}a-Burillo},
  {Garc{\'\i}a-Lorenzo}, {Gonz{\'a}lez-Mart{\'\i}n}, {Imanishi}, {Labiano},
  {Leist}, {Levenson}, {Ramos Almeida}, {Ricci}, {Rigopoulou}, {Rosario},
  {Stalevski}, {Ward}, {Esparza-Arredondo}, {Delaney}, {Fuller}, {Haidar},
  {H{\"o}nig}, {Izumi}, \& {Rouan}}]{2024ApJ...974..195Z}
{Zhang}, L., {Packham}, C., {Hicks}, E. K.~S., {et~al.} 2024, \apj, 974, 195

\end{thebibliography}
\bibliographystyle{aa}

\appendix

\section{Plot of the fitted parameters for a free PA and a fixed inclination angle}

\begin{figure}[ht]
    \centering
    \includegraphics[width=0.94\textwidth]{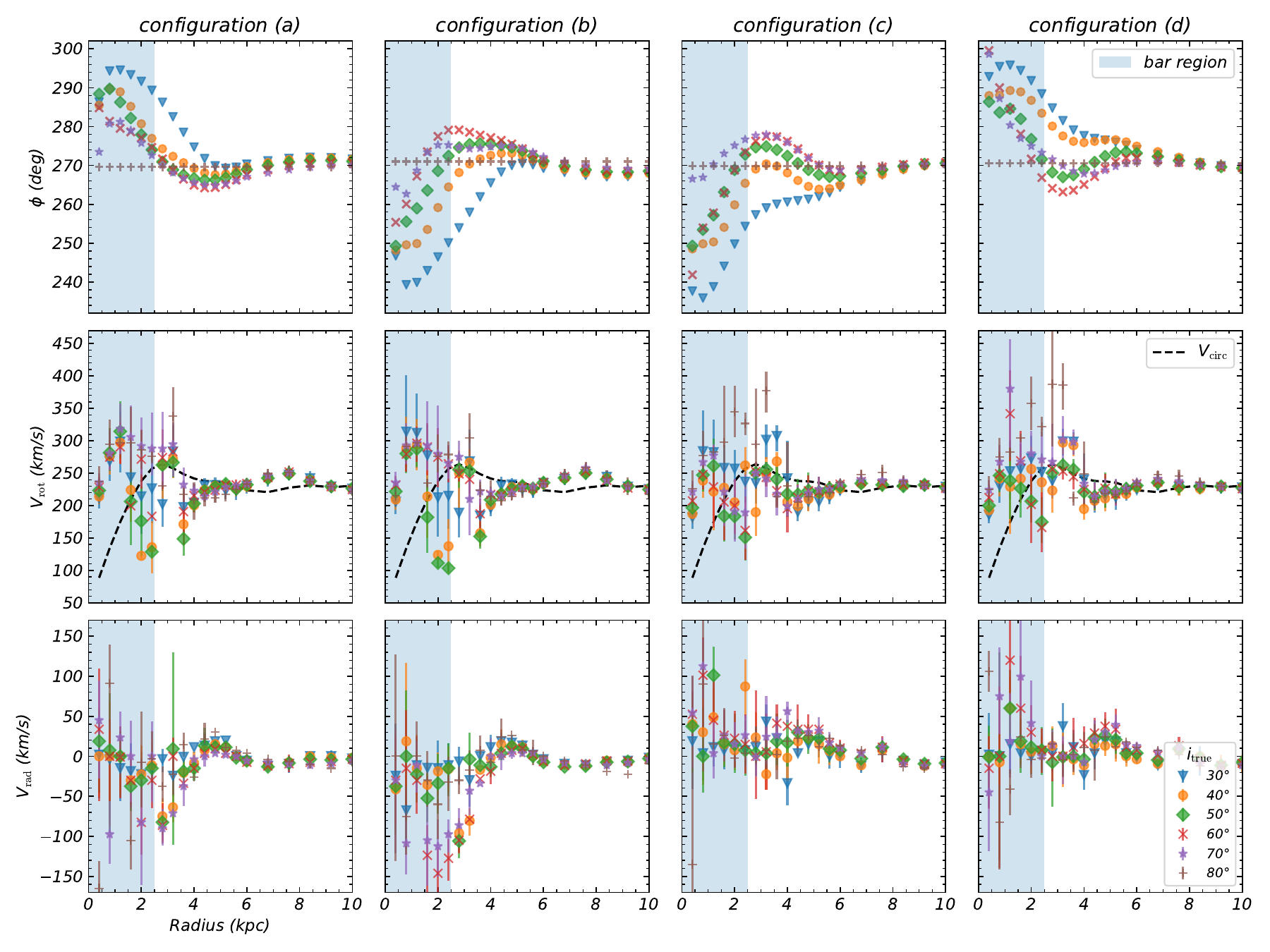}
    \caption{Same as Fig.~\ref{fig:fits_PA_free} but parameters were fitted with a free PA and a fixed inclination angle ($i=i_{\rm true})$.}
    \label{fig:fits_inc_fixed}
\end{figure}

\end{document}